\documentclass[a4paper,fleqn,usenatbib]{mnras}

\usepackage{newtxtext,newtxmath}
\usepackage[T1]{fontenc}
\usepackage{ae,aecompl}

\usepackage{graphicx}	% Including figure files
\usepackage{amsmath}	% Advanced maths commands
\usepackage{amssymb}	% Extra maths symbols

\title[S2CLS/EGS: Physical properties of faint SMGs]{The SCUBA-2 Cosmology Legacy Survey: The EGS deep field -- II.
Morphological transformation and multi-wavelength properties of faint submillimetre galaxies}
\author[J. A. Zavala et al.]{J. A. Zavala,$^{1, 2,3}$\thanks{E-mail: \href{mailto:jzavala@utexas.edu}{jzavala@utexas.edu}}
I. Aretxaga,$^{1}$
J. S. Dunlop,$^{2}$
M. J. Micha{\l}owski,$^{2}$
D. H. Hughes,$^{1}$
\newauthor
N. Bourne,$^{2}$ 
E. Chapin,$^{4}$
W. Cowley,$^{5}$
D. Farrah,$^{6}$
C. Lacey,$^{5}$
T. Targett,$^{7}$
\newauthor
P. van der Werf$^{8}$
\\ 
$^{1}$Instituto Nacional de Astrof\'{i}sica, \'{O}ptica y Electr\'{o}nica (INAOE), Luis Enrique Erro 1, Sta. Ma. Tonantzintla, 72840 Puebla, Mexico\\
$^{2}$Institute for Astronomy, University of Edinburgh, Royal Observatory, Blackford Hill, Edinburgh EH9 3HJ, UK\\
$^{3}$Department of Astronomy, The University of Texas at Austin, 2515 Speedway Boulevard Stop C1400, Austin, TX 78712, USA\\
$^{4}$Herzberg Astronomy and Astrophysics, National Research Council Canada, 5071 West Saanich Road, Victoria, BC V9E 2E7, Canada\\
$^{5}$Institute for Computational Cosmology, Department of Physics, University of Durham, South Road, Durham DH1 3LE, UK\\
$^{6}$Department of Physics, Virginia Tech, Blacksburg, VA 24061, USA\\
$^{7}$Department of Physics and Astronomy, Sonoma State University, 1801 East Cotati Avenue, Rohnert Park, CA 94928-3609, US\\
$^{8}$Leiden Observatory, Leiden University, PO Box 9513, 2300 RA Leiden, the Netherlands\\
}

\date{Accepted 2018 January 23. Received 2018 January 16; in original form 2017 April 25.}

\pubyear{2017}

\begin{document}
\label{firstpage}
\pagerange{\pageref{firstpage}--\pageref{lastpage}}
\maketitle

\begin{abstract}
We present a multi-wavelength analysis of galaxies selected at 450 and 850\! $\micron$ from the 
deepest SCUBA-2  observations in the Extended Groth Strip (EGS) field, which have an average depth of $\sigma_{\rm 450}=1.9$\! and  
$\sigma_{\rm 850}=0.46\rm\ mJy\ beam^{-1}$ over $\sim70$\ arcmin$^2$. 
The final sample comprises 95 sources: 56 (59\! \%) are detected at both wavelengths,
31 (33\! \%) are detected only at 850\! $\micron$, and 8 (8\! \%) are detected only at 450\! $\micron$. We identify  
counterparts for 75\! \% of the whole sample. The
redshift distributions of the 450 and 850\! $\micron$ samples peak at different redshifts with median values of $\bar{z}=1.66\pm0.18$ and $\bar{z}=2.30\pm0.20$, 
respectively. However,
the two populations have similar IR luminosities, SFRs, and stellar masses, with mean values of $1.5\pm0.2\times10^{12}\ L_\odot$,
$150\pm20\ M_\odot/{\rm yr}$, and $9.0\pm0.6\times10^{10}\ M_\odot$, respectively. This places most of our sources ($\gtrsim85$\! \%) on the high-mass
end of the `main-sequence' of star-forming galaxies. 
Exploring the IR excess vs UV-slope (IRX-$\beta$) relation we find that the most luminous galaxies are consistent with the Meurer law, 
while the less luminous galaxies lie below this relation. Using the results of a two-dimensional
modelling of the HST {\it H$_{\rm 160}$}-band imaging, we derive a median S\'ersic index of $n=1.4^{+0.3}_{-0.1}$ and a median half-light radius 
of $r_{1/2}=4.8\pm0.4$\! kpc. Based on a visual-like classification in the same band, we find that the
dominant component for most of the galaxies at all redshifts is a disk-like structure, although there is a transition from 
irregular disks to disks with a spheroidal component at $z\sim1.4$, which morphologically supports the 
scenario of SMGs as progenitors of massive elliptical galaxies.

\end{abstract}


%%%%%%%%%%%%%%%%%%%%%%%%%%%%%%%%%%%%%%%%%%%%%%%%%%

%%%%%%%%%%%%%%%%% BODY OF PAPER %%%%%%%%%%%%%%%%%%

\section{Introduction}\label{intro_secc}

Since their discovery, submillimeter-selected galaxies (hereafter SMGs) have revolutionized the field of galaxy formation and evolution. These sources 
were detected, for the first time, by the first submillimeter (850\! $\micron$) surveys taken with the James Clerk Maxwell Telescope (JCMT, \citealt{1997ApJ...490L...5S};
\citealt{1998Natur.394..248B}; \citealt{1998Natur.394..241H}), revealing that at least a fraction of the previously detected cosmic infrared background (CIB) by the 
space-based Cosmic Background Explorer (COBE, \citealt{1996A&A...308L...5P}; \citealt{1998ApJ...508..123F}) came from dust-enshrouded galaxies. These results
changed immediately our understanding of the cosmic star-formation history, and implied that surveys at both ultraviolet (UV)/optical and infrared (IR)/mm are necessary 
to completely understand it.

Thanks to the extensive follow-up studies carried out during the last two decades, we know that these are typically high-redshift galaxies ($\langle z\rangle\sim2-3$,
e.g. \citealt{2003MNRAS.342..759A,2007MNRAS.379.1571A};  \citealt{2005ApJ...622..772C}; \citealt{2012MNRAS.426.1845M}; \citealt{2012MNRAS.420..957Y}), with high star
formation rates (SFRs, $\ga300\ M_\odot\ {\rm yr}^{-1}$), large far-infrared  (FIR) luminosities ($\ga10^{12}\ {\rm L_\odot}$), large gas reservoirs ($\ga10^{10}\ M_\odot$), 
and orders of magnitude higher number density than local ultra-luminous infrared galaxies (see reviews by \citealt{2002PhR...369..111B} and \citealt{2014PhR...541...45C}). Furthermore, 
these galaxies are considered to be the progenitors of  massive elliptical galaxies (e.g. \citealt{1999ApJ...518..641L}; \citealt{2002MNRAS.331..495S}; 
\citealt{2014ApJ...788..125S}; \citealt{2014ApJ...782...68T}). For these reasons, this population is very important in our general comprehension of the stellar mass assembly, 
and therefore, in our understanding of the formation and evolution of galaxies over cosmic time. However, despite these significant efforts, our knowledge of
this population of galaxies is still not complete, since most of the samples come from single-dish telescope observations with large beam-sizes 
($\ga15$\! arcsec) at just one wavelength. This introduces several biases: (1) due to selection effects, observations at a single wavelength are not representative
of the whole population of galaxies (\citealt{2014MNRAS.443.2384Z}; \citealt{2013MNRAS.436.1919C}); (2) the poor angular resolution results in large position 
uncertainties leading to some misidentifications, as revealed by interferometric observations (\citealt{2013ApJ...768...91H}); (3) the high confusion noise caused by 
the large beam-sizes prevents from detecting galaxies with $L_{\rm FIR}\la10^{12}\ {\rm L_\odot}$ at high redshifts. 

Follow-up interferometric observations at submm bands exist now for more than a hundred SMGs (e.g.  \citealt{2006ApJ...640L...1I}; \citealt{2007ApJ...671.1531Y,2009ApJ...704..803Y};
\citealt{2011ApJ...726L..18W}; \citealt{2012ApJS..200...10S};  \citealt{2013ApJ...768...91H}; \citealt{2015ApJ...810..133I}; \citealt{2015A&A...577A..29M};
\citealt{2015ApJ...807..128S}; \citealt{2017A&A...608A..15B}), which alleviate the problem of positional uncertainty and source blending. Recent deep blank-field observations taken with
the Atacama Large Millimeter/submillimeter Array (ALMA) have allowed the detection of galaxies with SFRs $<100\ M_\odot\ {\rm yr}^{-1}$ 
(e.g. \citealt{2017MNRAS.466..861D}; \citealt{2016PASJ...68...36H}; \citealt{2017ApJ...835...98U}). These sources 
are also detected in small amplified samples towards clusters of galaxies thanks to gravitational amplification (e.g. \citealt{2017arXiv170304535P}). However, due to the 
small surveyed areas only a handful of galaxies are typically detected.
Additionally, these observations are taken at just one wavelength and may be not representative of the whole population. The achievement of wide-area surveys 
at different wavelengths which are necessary to solve these problems, as those planned with TolTEC\footnote{\url{http://toltec.astro.umass.edu}} on the LMT 
or in a small scale with ALMA, will take several years.
In the meanwhile, new deep single-dish telescope observations are coming along to minimize the biases in order to increase our knowledge of SMGs.

The SCUBA-2 Cosmology Legacy Survey (S2CLS; \citealt{2017MNRAS.465.1789G}) exploits the capabilities of the SCUBA-2 camera (\citealt{2013MNRAS.430.2513H}) 
on the JCMT, efficiently 
achieving large and deep (confusion limited) maps at both 450 and 850\! $\micron$ simultaneously, allowing us the detection of galaxies with 
$L_{\rm FIR}\la10^{12}\ {\rm L_\odot}$ up to $z\sim3$ (see \S\ref{lum_secc}). The higher angular resolution at 450\! $\micron$ results in a positional uncertainty of
$\sim1-2$\! arcsec, and therefore, in high accuracy associations than previous studies based on sample of galaxies selected from single-dish telescope observations.
Previous SCUBA-2 studies (including those from the S2CLS) have taken these advantages to characterize the physical properties of faint SMGs (\citealt{2013ApJ...776..131C}; \citealt{2013MNRAS.432...53G}; \citealt{2013MNRAS.436..430R};
\citealt{2016arXiv160500046H}; \citealt{2016MNRAS.458.4321K}; \citealt{2017ApJ...837..139C}; \citealt{2017MNRAS.469..492M}), although some of these studies have been focused on samples of either 450 or 850\! $\micron$-selected galaxies. 
Here we present a multi-wavelength counterpart analysis of a sample built from both 450 and 850\! $\micron$-selected galaxies, in order to minimize selection effects
and to investigate any differences in the physical properties of these galaxies.

This paper is organised as follows: sample selection and multi-wavelength data are described in \S2. The counterpart matching and identification process are reported in \S3. In \S4 we derive the physical
properties, such as redshifts, luminosities, SFRs, stellar masses, dust properties, and discuss the location of our galaxies in the star-forming main-sequence. The 
morphology classification as well as a possible morphological evolution is also discussed in \S4. Finally, our results are summarized in \S5.

All calculations assume a standard $\Lambda$ cold dark matter cosmology with $\Omega_\Lambda=0.68$, $\Omega_{\rm m}=0.32$, and
$H_0=67$ kms$^{-1}$Mpc$^{-1}$ (\citealt{2014A&A...571A..16P}).

\section{Multi-wavelength Data}

\subsection{S2CLS}\label{sample_selec}
The main data for this study comes from the deep 450 and 850\! $\micron$ observations taken with the SCUBA-2 camera in the EGS field 
as part of the S2CLS. The characteristics of observations 
and the data reduction process are described in the first paper of this
series (\citealt{2017MNRAS.464.3369Z}), where the source catalogues and number counts are also reported. 
These observations have a mean depth of $\sigma_{\rm 450}=1.9$ and $\sigma_{\rm 850}=0.46\rm\ mJy\ beam^{-1}$ (including instrumental and confusion noise) 
at 450 and 850\! $\micron$, respectively, within an area of $\sim 70$ arcmin$^2$. Along with other SCUBA-2 surveys (see \S\ref{intro_secc}),
these are some of the deepest observations taken
at these wavelengths with a single-dish telescope, enabling the detection of relatively faint galaxies (i.e. $0.6 \lesssim S_{850\micron}\lesssim 6\rm\ mJy$) that were unreachable by previous blank-field surveys. For comparison, the sources detected in the LABOCA ECDFS submillimetre Survey (LESS; \citealt{2009ApJ...707.1201W}), 
one of the largest and deepest contiguous map at submillimetre wavelengths before the achievement of the S2CLS, were limited to $S_{870\micron}\gtrsim 4\rm\ mJy$.  
Finally, the angular resolution at 450\! $\micron$ is $\theta_{\rm FWHM}\approx8$ arcsec and 
$\approx14.5$ arcsec at 850\! $\micron$.

\subsection{Ancillary data}

Thanks to the All-wavelength Extended Groth Strip International Survey (AEGIS\footnote{\url{http://aegis.ucolick.org/}}), this field has a 
panchromatic dataset from X-rays to radio wavelengths. We use these observations, among other catalogues described below, to identify the 
counterpart galaxies and to study their multi-wavelength properties. 

The radio and IR images are used as an intermediate step to associate each galaxy in our catalogue with an optical counterpart (see \S\ref{sec:identification}) 
For this purpose we use the VLA/EGS 20\! cm (1.4\! GHz) survey described by \citet{2007ApJ...660L..77I}. These observations have an angular resolution of 
FWHM $\approx3.8$\! arcsec with a $5\sigma$ detection limit of 50\! $\mu{\rm Jy}$ in the deepest region. As IR constraints, we use the catalogue derived
from {\it Spitzer}/MIPS observations at 24\! $\micron$ which are part of the Far-Infrared Deep Extragalactic Legacy Survey 
(FIDEL\footnote{\url{http://irsa.ipac.caltech.edu/data/SPITZER/FIDEL/}}). The $3\sigma$ detection
limit is 30\! $\mu{\rm Jy}$ and it has an angular resolution of 5.9\! arcsec. In addition, the 8\! $\micron$ catalogue derived using {\it Spitzer}/IRAC observations
(\citealt{2011ApJS..193...13B}) is also exploited for this purpose, with a 5$\sigma$ limit of 22.3\! mag and an angular resolution of 2.2\! arcsec.

We use the multi-wavelength catalogues compiled by the 3D-HST team (\citealt{2012ApJS..200...13B}; \citealt{2014ApJS..214...24S}; \citealt{2016ApJS..225...27M}) 
to finally associate our radio or IR counterpart with an optical galaxy. These catalogues are built mainly using {\it H}-band selected galaxies from 
{\it HST}(F160W) imaging with a median $5\sigma$ depth of 26.4\! mag (AB) and include photometry from the {\it u}-band to 8\! $\micron$. 
We also use an {\it Spitzer/}IRAC 3.6$+$4.5\! $\micron$ selected catalogue (\citealt{2011ApJS..193...13B}; \citealt{2011ApJS..193...30B}) which contains photometry from 
the ultraviolet to 70\! $\micron$. 
The catalogue includes sources up to $5\sigma$ of 23.9\! mag at 3.6\! $\micron$. As discussed in \S\ref{z_dist_secc} and \S\ref{main_seq_secc} these catalogues
include redshift information and stellar-population parameters, which are used to understand the physical properties of our sample.

To investigate the AGN contamination in our sample, the catalogue of the AEGIS-X Deep (AEGIS-XD) survey\footnote{\url{http://www.mpe.mpg.de/XraySurveys/AEGIS-X/}} 
(\citealt{2015ApJS..220...10N}) is used. This program combines deep {\it Chandra} observations in the central region of EGS with previous {\it Chandra} observations 
of a wider area (AEGIS-X Wide, \citealt{2009ApJS..180..102L}), achieving a total nominal exposure depth of 800ks in the central region. This makes this program one 
of the deepest X-ray survey in existence. 

To extract far-infrared photometry we use {\it Herschel} observations obtained with the Photodetector Array Camera and Spectrometer 
(PACS, \citealt{2010A&A...518L...2P}) and the Spectral and Photometric Imaging Receiver (SPIRE, \citealt{2010A&A...518L...3G}) instruments, which are part of the PACS 
Evolutionary Probe (PEP, \citealt{2011A&A...532A..90L}) and the {\it Herschel} Multi-tiered Extragalactic Survey (HerMES, \citealt{2012MNRAS.424.1614O}) programs.
 We obtained the {\it Herschel} fluxes of each SCUBA-2 source as in \citet{2017MNRAS.469..492M} in the following way.
We extracted 120 arcsec wide stamps from all five {\it Herschel} maps around the position of each SCUBA-2 source.
Then we processed the PACS ($100$ and $160\,\mu$m) maps  by simultaneously fitting Gaussian functions with
the FWHM of the respective resolution of the maps, centred at the positions of all  $24\ \micron$ sources located
within these cut-outs, and at the positions of the SCUBA-2 IDs. 
Then, to deconvolve the SPIRE ($250$, $350$ and $500\,\mu$m)
maps in a similar way, we used the positions of the $24\ \micron$ sources detected with PACS ($\ge3\sigma$),
the positions of all SCUBA-2 ID positions (or the submm positions if no
radio or mid-IR ID had been secured). 
The errors were computed from the covariance matrix of the fit, in which the free parameters are simply the heights of the Gaussian
beams fitted at each input position. Then the confusion noise of 5.8, 6.3 and 6.8\,mJy\,beam$^{-1}$ at 250, 350 and 500\,\micron,
respectively \citep{nguyen10} was added in quadrature. The fitting was performed using the IDL
{\sc Mpfit}\footnote{\url{purl.com/net/mpfit}} package \citep{mpfit}.

\section{Source identification}

For this study, we limit our sample to those sources detected at $>3.75\sigma$ at $450$ or $850\ \micron$. At this threshold, the contamination due to false-detections is 
expected to be $\la 5$ per cent (\citealt{2017MNRAS.464.3369Z}), which is acceptable for the goals of this paper. On the other hand, sources detected at two (or more) bands have 
a higher effective S/N, for example, a source detected at $3\sigma$ at two different wavelengths has an effective S/N of $\sim4\sigma$. For 
this reason, we also include all the sources detected at both wavelengths with S/N $>3.0$.  A source is considered detected at both wavelengths 
if the separation between the 450 and 850\! $\micron$ position is $\la7$ arcsec, which corresponds to $2.5\sigma$ the joined expected  positional uncertainty 
added in quadrature according to the beam-size and S/N (\citealt{2007MNRAS.380..199I}).

Following these criteria, our final sample consist of 95 sources: 56 (corresping to 59\! \% of the total sample) are detected at both wavelengths, 
31 (33\! \%) are detected only at 850\! $\micron$, and 8 (8\! \%) are detected only at 450\! $\micron$.

\subsection{Counterpart matching}\label{sec:identification}
Traditionally, the first step to find a SMG counterpart is to associate it with a radio or IR source, since the source density at optical wavelengths
is much higher and therefore many potential counterparts lie within a typical  search radius.

The radio band (1.4\! GHz) is chosen because it also traces 
recent star formation via synchrotron radiation and there is a well-known FIR-radio correlation (\citealt{2002ApJ...568...88Y}; \citealt{2014ApJ...784....9B}). 
On the other hand, 24\! $\micron$ observations are sensitive to the warm dust emission and SMGs have been found to be bright at this wavelength
(e.g. \citealt{2006MNRAS.370.1185P}). 
Additionally, the surface density of sources is relatively low at these bands and consequently the probability for a misidentification is expected to be low, 
although some ALMA studies reported a misidentification fraction of up to  $\sim30$ per cent (\citealt{2013ApJ...768...91H}). 
Over the last decade, observations at 8\! $\micron$ have also been used for counterpart identifications (e.g. \citealt{2006ApJ...644..778A}; \citealt{2006MNRAS.370.1185P};
\citealt{2011MNRAS.413.2314B}; \citealt{2012MNRAS.426.1845M}; \citealt{2012MNRAS.420..957Y}) since these trace the emission from the 
older and mass-dominant stellar populations in this kind of galaxies and, although the surface density is higher,  these observations are usually
deeper than those at 24\! $\micron$, increasing the fraction of identifications.

 Using these three wavebands independently, we search for counterparts with a variable search radius equal to $2.5\sigma$ (corresponding to 
 $\sim96$ per cent probability), where $\sigma$ is the positional uncertainty of each galaxy based on its S/N and the FWHM of the beam, 
 as described by \citet{2007MNRAS.380..199I}. If available, the 450\! $\micron$ position is favored over the 850\! $\micron$. Additionally, we imposed a minimum search radius of 4 arcsec to account for systematic astrometry differences between catalogues and for the positional uncertainties of the
 NIR/radio sources. Furthermore, this minimum search radius help us to increase 
the fraction of identifications,  since it has been found that some SMGs detected with relatively high S/N lie outside of their nominal $2.5\sigma$ positional uncertainties 
(\citealt{2013ApJ...768...91H}). 
Then, if a potential counterpart is found, we estimate the statistical significance of the association or in other words, 
the corrected Poisson probability, $p$, that the counterpart candidate has been selected by chance, following the  method described by \citet{1986MNRAS.218...31D}
(see also \citealt{1989MNRAS.238.1171D} and \citealt{2007MNRAS.380..199I}).

An empirical value of $p<0.05-0.1$ to define robust counterparts for SMGs is commonly adopted in the literature (e.g. \citealt{2006MNRAS.370.1185P}; 
\citealt{2009MNRAS.398.1793C}; \citealt{2012MNRAS.420..957Y}),
however, using ALMA observations \citet{2016ApJ...820...82C} showed that there is no statistical difference between the accuracy if we adopt $p<0.05$ 
or $p<0.1$ (i.e. consistent within the error bars) but a better completeness if we select the later. Therefore, we adopt as probable counterparts 
those galaxies with $p<0.1$. For the case of 
galaxies with multiple candidates (20 out of 71), we select the one with the lowest $p$-value, following \citet{2017MNRAS.469..492M}, who have shown that this procedure
recovers the galaxy with the dominant contribution to the submm flux density in most cases ($\sim86$\! \%).

Using these criteria, and after rejecting six candidates based on their discrepant FIR photometric redshifts (see \S\ref{z_dist_secc}), we achieve a successful 
identification rate of $\sim75$\! \% (71 out of 95) for the whole sample. In terms of the SCUBA-2-band detections, the successful identification rate 
is 82\! \%  for those galaxies detected at both wavelengths, 88\! \% for those detected only at 450\! $\micron$, and 62\! \% for those detected 
only at 850\! $\micron$. The lower detection rate for the 850\! $\micron$-only detected galaxies may reflect the higher redshift nature of these galaxies (see \S\ref{z_dist_secc}),
albeit the larger beam-size and the source blending translate into a larger positional uncertainty, which can also explain the lower fraction of counterparts.

Once we have the radio/IR counterparts, we match these sources with the 3D-HST catalogue (\citealt{2014ApJS..214...24S}; \citealt{2016ApJS..225...27M}) 
using a search radius of 1.5 arcsec (although all 
the matches lie within 1 arcsec). For those galaxies which lie outside of the 3D-HST coverage (13 out of 71), we use instead the IRAC catalogue 
(\citealt{2011ApJS..193...13B, 2011ApJS..193...30B}). All the 71 galaxies with radio/IR counterparts also have an optical counterpart in these catalogues. 
All properties are summarized in Table \ref{catalogue}.

\subsection{Reliability of galaxy identifications}

Before deriving the physical properties of the sample, it is important to test the reliability of our identification methodology and to quantify the misidentification 
fraction. It is clear that submm/mm interferometry represents the best way of identifying counterparts correctly (e.g. 
\citealt{2007ApJ...671.1531Y,2009ApJ...704..803Y}; \citealt{2011ApJ...726L..18W}; \citealt{2012ApJ...761...89B}; \citealt{2012A&A...548A...4S}; \citealt{2013ApJ...768...91H}). However, the 
archival submm/mm interferometry data is scarce in this field (which is, furthermore, inaccessible to ALMA). From the PHIBSS survey (\citealt{2013ApJ...768...74T}), we found
6 fields targetted with  IRAM Plateau de Bure millimeter interferometer (PdBI) whose detections lie close ($<10$\ arcsec) to sources in our catalogue. 
These observations were designed to map the $^{12}\rm CO(3-2)$ transition in massive ($M_*>2.5\times10^{10}\rm\ M_\odot$), main-sequence star-forming galaxies at 
$z\sim1.2$ and $2.2$. Since it has been shown that SMGs are bright in CO emission (e.g. \citealt{2005MNRAS.359.1165G}) and giving the properties of the PHIBSS 
targetted galaxies, we can assume that the CO emission is associated with an SMG. Moreover, the probability of finding by chance a CO line emission close to our SMGs is low, 
based on the results from blind CO surveys (\citealt{2016ApJ...833...67W}). Assuming then that these six $^{12}\rm CO(3-2)$ emission lines 
come from the SMGs, we can test our galaxy identifications. We find that for 5 galaxies we have identified the correct counterpart with our method. The remaining source is one of those with only 850\! $\micron$ detection, which has a larger positional uncertainty. Alternatively, the CO line might have been too faint to be detected by PHIBSS. Based on these results, we estimate 
a correct counterpart accuracy of $\gtrsim83$\ \%.

On the other hand, \citet{2013MNRAS.436.1919C} estimated the counterpart contamination when using the SCUBA-2 450\ $\micron$ position based on the reduced beamsize with 
respect to 850\ $\micron$, assuming a reduction in counterpart contamination proportional to the reduction in sky area searched for potential counterparts, and adopting
a $\sim30$\ \% of incorrect associations for the 850\ $\micron$ sources (\citealt{2013ApJ...768...91H}). Based on this, they inferred a $\sim5$\ \% of misidentifications 
for 450\ $\micron$ sources. Using these values, and considering that only $33$\ \% of our sample lack 450\ $\micron$ positions, we estimate that $\sim13$\ \% of the sample 
could have incorrect identifications. 

\begin{figure}
  \includegraphics[width=90mm]{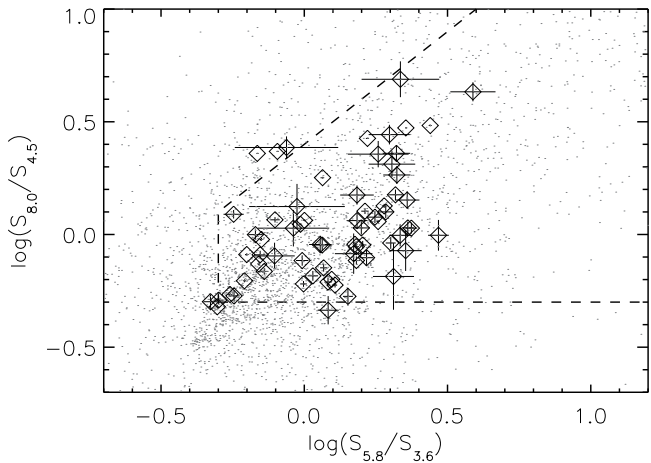}
  \caption{ {\it Spitzer}/IRAC $S_{5.8_{\mu\rm m}}/S_{3.6_{\mu\rm m}}$ vs. $S_{8_{\mu\rm m}}/S_{4.5_{\mu\rm m}}$ colour-colour diagram for the associated counterparts
  to our SMGs (black diamonds). The dashed lines enclose the IRAC colour cuts suggested for SMG counterpart identification by \citet{2008MNRAS.389..333Y}, 
  where most of our identified SMGs lies. The grey dots represent other galaxies in the 3D-HST catalogue in the same field. }
  \label{irac_colors}
\end{figure}

Finally, we compare the IRAC colours of our identified galaxies with the IRAC colour-colour space diagram proposed for SMG counterpart
identification by \citet{2008MNRAS.389..333Y}. Securely identified SMGs (with radio or submm interferometric observations) are known to lie within this proposed diagram 
(\citealt{2013MNRAS.431..194A}), and may help us to inquire into the reliability of our identifications. As shown in Fig. \ref{irac_colors}, most of our identified 
counterparts ($\sim95$\ \%) lie on this space (and the remaining sources lie very close to the edges), which further supports the associations. 

In summary, based on these three independent tests, we infer that our identification method 
recovers the correct counterpart with an accuracy of $\gtrsim 85$\ \%, which is acceptable for an statistical characterization of the
population.

% PHIBSS: min-> 0.12 Jy Km s;    Area ASPECS: 4,400 arcsec^2 = 1.23 arcmin^2;     1 source (max 3) per 75" pointing

\section{Derived properties}
\subsection{Redshift distribution}\label{z_dist_secc}

The optical catalogues described above include redshift estimations for most of the sources. For those galaxies with identified counterparts in the 3D-HST catalogue, 
we use the combined {\sc `Z-BEST'} redshift information (hereafter $z_{\rm opt}$). The best redshift is: (1) spectroscopic long-slit redshift, if available in the compilation by 
\citet{2014ApJS..214...24S}; 
(2) spectroscopic {\it HST} grism redshift (\citealt{2016ApJS..225...27M}), if there is no archive spectroscopic redshift and if the grism spectrum is not flagged as faulty;
or  (3) optical photometric redshift derived from the {\sc EAZY} code (\citealt{2008ApJ...686.1503B}). 
For the small fraction of sources with no 3D-HST counterpart, we use the photometric 
redshifts reported in the IRAC catalogue (\citealt{2011ApJS..193...30B}). Although these redshifts have been estimated using the {\it Rainbow} code 
(\citealt{2011ApJS..193...13B}), the authors have shown
that there is a good agreement with the values estimated using {\sc EAZY}. The typical uncertainty for the grism-based redshifts is 
$\Delta z/(1+z)\approx0.003$ (\citealt{2016ApJS..225...27M}), and $\Delta z/(1+z)\la0.04$ for the photometric redshifts derived from {\sc EAZY} or {\it Rainbow} 
(\citealt{2011ApJS..193...30B}; \citealt{2014ApJS..214...24S}) with a catastrophic failure rate of less than 10 per cent, when considering only $z>0.5$ sources (although, most of them are optically-selected galaxies). However, when considering only the galaxies in our catalogue,  we find a median uncertainty of $\Delta z/(1+z)=0.05$ for the photometric redshifts.

\begin{figure}
  \includegraphics[width=90mm]{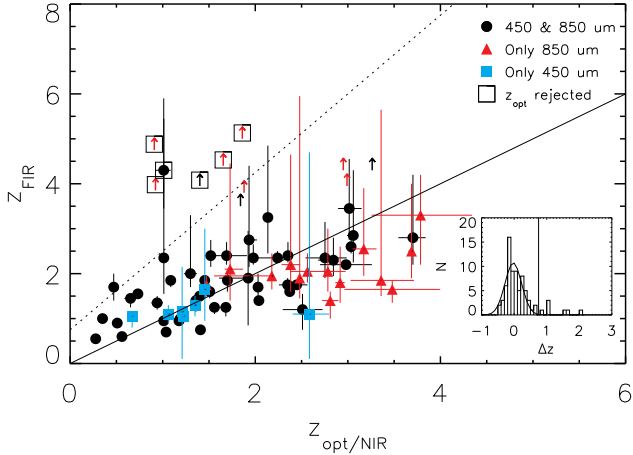}
  \caption{Comparison between  FIR photometric redshifts, derived from {\it Herschel} and SCUBA-2 observations ($z_{\rm{FIR}}$), and optical-infrared 
  spectroscopic or photometric redshifts ($z_{\rm{opt}}$) of the counterparts (see \S \ref{z_dist_secc}). Those sources with only lower limits are 
  marked with upward arrows. The inset plot shows a histogram of the relative 
  difference ($\Delta z=(z_{\rm{FIR}}-z_{\rm{opt}})/(1+z_{\rm{opt}})$) between the two estimations, which is well fitted with a Gaussian distribution. 
  The sources with a discrepant redshift above 3$\sigma$ of the Gaussian distribution (represented by the dotted line) are rejected given the potential of
  misidentification (see \S\ref{z_dist_secc}). The colours indicate if the source is detected at both wavelengths (black symbols) or only in a single band (blue and red
  symbols for 450 or 850\! $\micron$, respectively). }
  \label{redshift_compa}
\end{figure}

On the other hand, we estimate photometric redshifts for the whole sample using the rest-frame FIR photometric data 
provided by {\it Herschel} 100, 160, 250, 350, 500\! $\micron$, and our SCUBA-2 observations at 450 and 850\! $\micron$ 
(hereafter $z_{\rm FIR}$). Though this method is not as accurate as the optical photometric redshifts, it allows us to estimate
the redshift for sources with no counterparts, and therefore, to derive a complete redshift distribution.
We fit an average SMG SED template (\citealt{2010A&A...514A..67M}) to the flux density measured at these wavelengths. 
This is the average SED of 70 SMGs with spectroscopic redshifts, where GRASIL models (\citealt{1998ApJ...509..103S}) were fitted to the photometry and then averaged.
During this procedure the adopted flux calibration errors are 2 \& 4\% at 100 and 
160\ $\micron$\footnote{\url{http://herschel.esac.esa.int/twiki/pub/Public/PacsCali} \url{brationWeb/pacs_bolo_fluxcal_report_v1.pdf}}, 
4\% for the SPIRE bands (\citealt{2013MNRAS.433.3062B}), and 10 \& 5\% for the SCUBA-2 450 and 850\ $\micron$ (\citealt{2013MNRAS.430.2534D}).
In case of non-detections, we incorporate the upper limits into the  fitting through a
survival analysis (\citealt{1986ApJ...306..490I}) following the same procedure as in \citet{2007MNRAS.379.1571A}. For some sources only lower limits 
on redshifts could be derived (identified with upward arrows in Fig. \ref{redshift_compa}) due to the high number of non-detections. 
These $z_{\rm FIR}$ values are compiled in Table \ref{catalogue}.

 In order to investigate the reliability of these photometric redshifts, 
we compare the $z_{\rm FIR}$ with the optical redshifts described above for those galaxies with optical counterparts. As can be seen in Fig. \ref{redshift_compa},
there is generally a good agreement between both estimations within the large scatter. The relative difference between these methods
($\Delta z=(z_{\rm{FIR}}-z_{\rm{opt}})/(1+z_{\rm{opt}})$)
is well fitted with a Gaussian function (see inset plot in Fig. \ref{redshift_compa}) with a standard deviation of 0.25. However, there are six sources (marked with 
squares in the figure), for which the two redshift estimations differ dramatically ($z_{\rm FIR}=4-5$ vs. $z_{\rm opt}\sim1$). This may reflect an optical misidentification 
due to a nearby galaxy in the line of sight (e.g. \citealt{2014MNRAS.444.1884B}) or because the identified optical galaxy is in fact lensing a more distant submm source 
(e.g. \citealt{2015MNRAS.452..502G}).
All of these galaxies have S/N\ $>5$ and a low probability of being a false detection.  For these reasons,
we reject these  optical associations (hence also their $z_{\rm opt}$) for the rest of the analysis, in order to avoid introducing any bias (following 
\citealt{2016MNRAS.458.4321K} and \citealt{2017MNRAS.469..492M}).

\begin{figure*}
  \includegraphics[width=170mm]{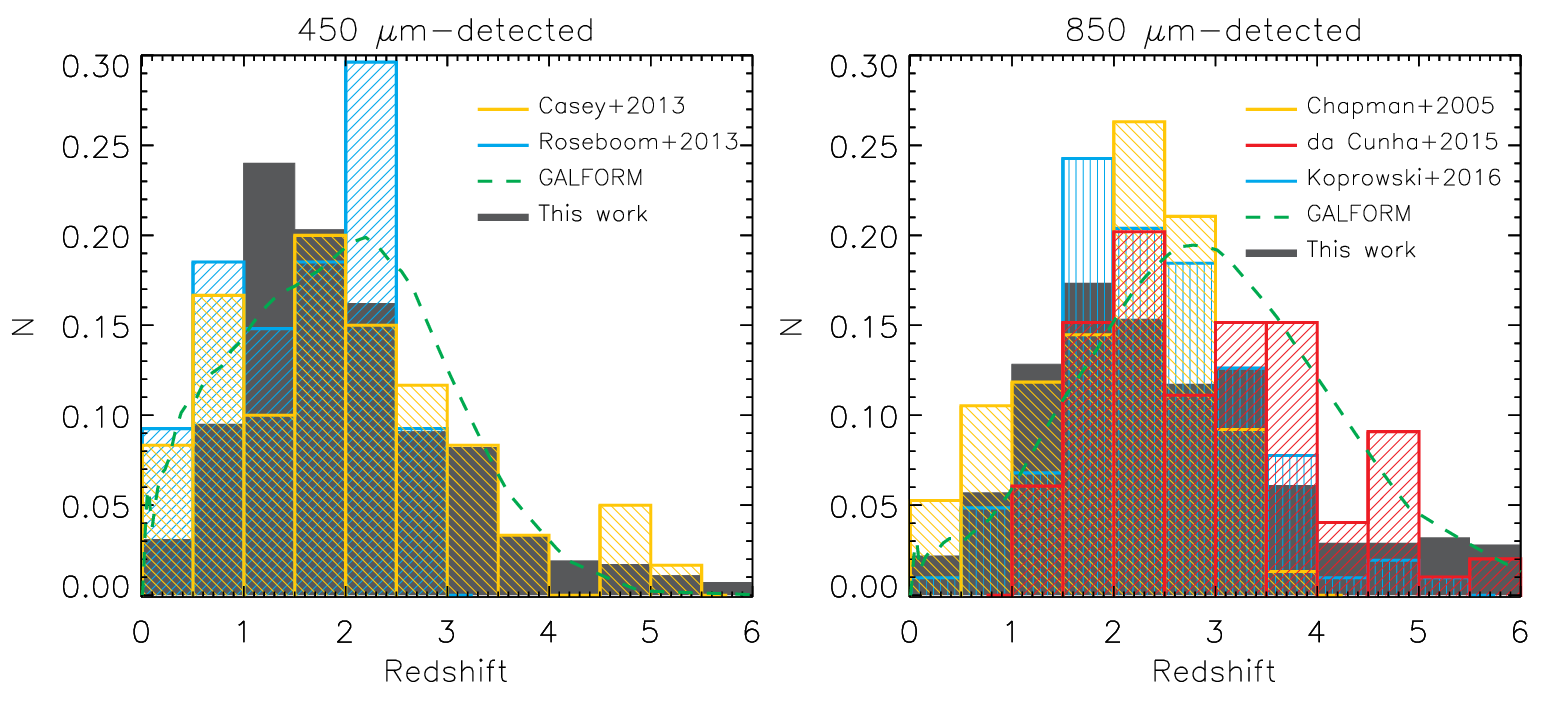}
  \caption{The normalized redshift distributions of our 450 and 850\! $\micron$-detected sources ({\it left} and {\it right}, respectively) are 
  represented by the grey histograms. The 450\! $\micron$ distribution has a median redshift of $z_{\rm med}=1.66\pm0.18$ and the 850\! $\micron$ one a median
  of $z_{\rm med}=2.30\pm0.20$. For comparison, the results from previous observational studies are plotted in each panel (\citealt{2013MNRAS.436..430R}; \citealt{2005ApJ...622..772C}; 
  \citealt{2013MNRAS.436.1919C}; \citealt{2015ApJ...806..110D}; \citealt{2016MNRAS.458.4321K}), as well as the theoretical predictions of the {\sc GALFORM} model 
  (\citealt{2016MNRAS.462.3854L}) matched to the mean S2CLS/EGS depth.  }
  \label{z_dist}
\end{figure*}

A trend can be seen in Fig. \ref{redshift_compa}, where our $z_{\rm FIR}$ is systematically lower than the optical redshift at $z_{\rm opt}>2$. 
This may reflect the fact that our SED template is represented by a single temperature and reveals a relation between redshift and dust temperature, 
where warmer temperatures are necessary at higher redshifts to correct for this effect. 
The same trend was actually found in the photometric redshift  
of red {\it Herschel} sources (\citealt{2016ApJ...832...78I}). 
However, due to the small number of photometric measurements for some sources in our catalogue and to the large scatter of this trend, we decide to keep our single 
dust temperature template for this purpose. As described below, these results are only used to complete the redshift distribution of all the sources in our catalogue. The rest of the analyses in the paper are focused only on those galaxies with optical counterparts. A detailed discussion of the dust temperature evolution is presented in \S\ref{dust_temp_secc}.

The final redshift distributions for all the 450 and 850\! $\micron$-selected galaxies are shown in Fig. \ref{z_dist}. These include optical spectroscopic redshifts 
(36 sources, including {\it HST} grism-derived redshifts), optical photometric redshifts (35 galaxies), and FIR redshifts for those sources with no counterparts (24 sources).
Individual values are reported in Table \ref{catalogue}. 
In order to take into account the uncertainties in our values, the redshift distributions 
have been estimated by stacking the redshift likelihood distribution of each source. The stacked redshift distributions expand between $0<z<6$, although, as expected 
due to the $k$-correction at NIR and radio wavelengths, all the galaxies with identified counterparts lie at $z<4$. The median redshifts of the
distributions are $z_{\rm med}=$ $1.66\pm0.18$ and $2.30\pm0.20$ (with the errors derived from a bootstrap method) for the 450 and 850\! $\micron$-selected galaxies, respectively.
This supports the statement that galaxies selected at different wavelengths have different redshift distributions as previously reported (e.g. \citealt{2014MNRAS.443.2384Z};
\citealt{2015A&A...576L...9B}). Actually, all (but one) of the galaxies detected only at 450\! $\micron$ (i.e. with no 850\! $\micron$ detection) lie at $z\la1.5$, 
while all the galaxies detected only at 850\! $\micron$ lie at $z\ga1.5$. As described in the next sections, this is the main difference between the galaxies
detected at only one of the two wavebands.

\subsubsection{Comparison with previous surveys}

We compare our redshift distributions with those derived from previous similar studies in Figure \ref{z_dist}.

\citet{2013MNRAS.436..430R} studied a sample of 450\ $\micron$-selected galaxies from the S2CLS observations in the UDS and COSMOS fields with a depth similar to this work, 
although with a S/N threshold of 4. Their  median redshift of $1.4\pm0.2$ is close to our value of $1.66\pm0.18$, however, as it can be seen in the figure, they have no sources with $z>3$. 
This is expected since their redshift distribution is limited to those galaxies with optical counterparts, and therefore, the highest redshift galaxies are missing. On the 
other hand, \citet{2013MNRAS.436.1919C} presented an analysis for 450 and 850\ $\micron$ galaxies on the COSMOS field covered with SCUBA-2 observations. 
For the 450\ $\micron$ sample they reported a median 
redshift of $z=1.95\pm0.19$. This value is slightly higher than the one found in this work, however, their sample comes from larger but shallower observations selecting, therefore, brighter
galaxies that are not present in our smaller field (i.e. sources with $S_{450\mu\rm m}>20\rm\ mJy$). This is consistent with the picture that galaxies with higher fluxes are located preferentially at higher redshifts as suggested by previous studies 
(e.g. \citealt{2005MNRAS.358..149P}; \citealt{2014MNRAS.444..117K}). Finally, we also compare our 450\ $\micron$ redshift distribution with theoretical predictions of the
{\sc GALFORM} semi-analytical model (\citealt{2016MNRAS.462.3854L}) for galaxies with $S_{450}>7$\ mJy.  This flux density 
limit corresponds to the average limit of our sample (S/N\ $>3.75$).
% , although we have some sources with lower flux densities (since we are selecting  galaxies
% with $3<\rm S/N<3.75$ that are also detected at 850\ $\micron$, see \S\ref{sample_selec}). 
The model predicts a median redshift of $1.97$, which is consistent with our
value within $\sim1.7\sigma$. This model also includes the rare bright galaxies that our relatively small map cannot constrain, which could account for the 
mild difference between the redshift distributions. Actually, the predictions of the model are in better 
agreement with the results of \citet{2013MNRAS.436.1919C}, where, as mentioned before, the brightest galaxies are better sampled due to the larger area mapped. 
%  with Kolmog\'orov-Smirnov probabilities of 0.5 vs 0.25.

At 850\ $\micron$,  \citet{2013MNRAS.436.1919C} reported a median redshift of $=2.16\pm0.11$ which is just below but consistent with our median value of $2.3\pm0.2$. Nevertheless,
our distribution shows a flat high-redshift tail between $4<z<6$ that is not present in the \citeauthor{2013MNRAS.436.1919C} result. This is not  surprising 
since their redshift distribution is limited to those galaxies with  optical counterparts. On the other hand, \citet{2015ApJ...806..110D} presented 
the redshift distribution of sources with 870\ $\micron$ fluxes above 4\ mJy (a factor of $\sim2$ larger than our threshold) derived from {\sc magphys} SED modeling, which has a 
median redshift of $2.7\pm0.1$. In contrast to our distribution, they have no sources at $z<1$ (Fig. \ref{z_dist}.). This may reflect the fact that we are using the 450\ $\micron$ 
information to select faint sources below our formal detection limit, increasing the completeness of our sample. As in this work,  \citet{2016MNRAS.458.4321K} used deep
S2CLS observations to study the properties of 850\ $\micron$ galaxies in the COSMOS field. Their redshift distribution with a median of $2.38$ is in very good agreement
with our result, as can be  seen from Figure \ref{z_dist}. On the theoretical side, the {\sc GALFORM} model predicts a median of $2.88$ for galaxies with 
$S_{850}>1.5$\ mJy. This value is higher than ours but consistent within $3\sigma$.

In general,
our results are in broad agreement with previous studies and show a trend where longer wavelength observations select, on average, higher redshift galaxies. 

\subsection{IR luminosities, SFRs, \& dust properties}\label{dust_prop_secc}

The rest-frame FIR photometry allows us to derive infrared luminosities (which can be converted to star formation rates) and dust temperatures once a SED is fitted to the data.
We use a modified blackbody function which is described by
\begin{equation}
  S_\nu\propto \{1- \exp[-(\nu/\nu_0)^\beta]\}B(\nu,T_{\rm d}),
\label{eq:mbb}
\end{equation}
where $S_\nu$ is the flux density at frequency $\nu$, $\nu_0$ is the rest-frame frequency at which the emission becomes optically thick,
$T_{\rm d}$ is the dust temperature,  $\beta$ is the emissivity index, and $B(\nu,T_{\rm d})$ is the Planck function at 
temperature $T_d$. To minimize the number of free parameters,
the emissivity index, $\beta=1.6$, is fixed (previous observational works suggest $\beta=1.5-2$; e.g. \citealt{2001MNRAS.327..697D}; \citealt{2003MNRAS.343..585F}; \citealt{2009MNRAS.398.1793C};
\citealt{2012A&A...539A.155M}), as well as $\nu_0=c/100\ \micron$ (\citealt{2013Natur.496..329R}; \citealt{2016arXiv161103084S}), where $c$ is the speed of light.
Furthermore, in order to break the temperature-redshift degeneracy
(e.g. \citealt{2002PhR...369..111B}), we fix the modified blackbody at the redshift provided by the optical catalogues (see \S \ref{z_dist_secc}). Therefore, 
galaxies with no counterparts (24/95) are not included in this analysis. All the derived properties, named luminosities, SFRs, and dust temperatures, 
are reported in Table \ref{catalogue}, and are discussed below. If we choose instead $\beta=2.0$, the derived IR luminosity does not change (within 1\%) but the dust
temperature decreases on average by $\sim10$\%.

\subsubsection{Luminosity}\label{lum_secc}

In Fig. \ref{z_SFR} we plot the IR luminosity ($8-1000\ \micron$) as a function of redshift for our sources with optical counterparts, and the detection limit (taking the best 
case between the 450 and 850$\ \micron$ limit at each redshift) assuming a modified blackbody with $T_{\rm d}=47\ K$ (the average temperature of the sample)
and $T_{\rm d}$ increasing with redshift (as measured 
in our data, see \S\ref{dust_temp_secc}). As it can be seen, the  detection limit predicted by the fixed dust temperature SED does not reproduce our estimations, over-predicting 
the $L_{\rm IR}$ limit at low redshift and underpredicting it at high redshift. This effect was also noticed by \citet{2016ApJ...832...78I}. On the other hand, the expected detection 
limit for a modified blackbody with dust temperature increasing with redshift is in much better agreement with our measurements. This provides additional evidence of a relation 
between $T_{\rm d}-z$ (see \S\ref{z_dist_secc} and \S\ref{dust_temp_secc}). 

Our survey is deep enough to detect galaxies down to an IR luminosity of $L_{\rm IR}\sim1.5\times10^{12}\ L_\odot$ at any redshift below $z\sim4$ (see Fig. \ref{z_SFR}), 
which corresponds to a SFR of $150\ M_\odot\ {\rm yr}^{-1}$, assuming the \citealt{1998ARA&A..36..189K} relation for a \citealt{2003PASP..115..763C} IMF. 
This highligths the depth of our survey, which actually allows us to detect several galaxies with $20<{\rm SFR}<100$ at $1<z<2$  ({as other recent deep S2CLS studies, e.g.
\citealt{2016MNRAS.458.4321K}; \citealt{2017MNRAS.467.1360B}). This population of galaxies was
unreachable by previous submillimeter surveys with single-dish telescopes in blank fields. 
Also shown in Fig. \ref{z_SFR} is the evolution of the knee of the IR luminosity function ($L^*$, 
\citealt{2013MNRAS.432...23G}), which is slightly fainter (less than a factor of $1.5$) than our sensitivity limit.

The mean IR luminosity of our sample is $1.5\pm0.2\times10^{12}\ L_\odot$ which corresponds to a mean SFR of $150\ M_\odot\ {\rm yr}^{-1}$ 
(where the errors have been estimated by a bootstrapping method), covering 
a range from $L_{\rm IR}=0.3\times10^{11}-7.4\times10^{12}\ L_\odot$ (SFR=$3-740\ M_\odot\ {\rm yr}^{-1}$).
We notice that at a fixed redshift, the galaxies detected at both wavelengths (black circles in Fig. \ref{z_SFR}) are, on average, more luminous than those detected at
only one wavelength (blue and red circles for 450 and 850 \! $\micron$-only detected sources, respectively), as expected.

\begin{figure}
  \includegraphics[width=90mm]{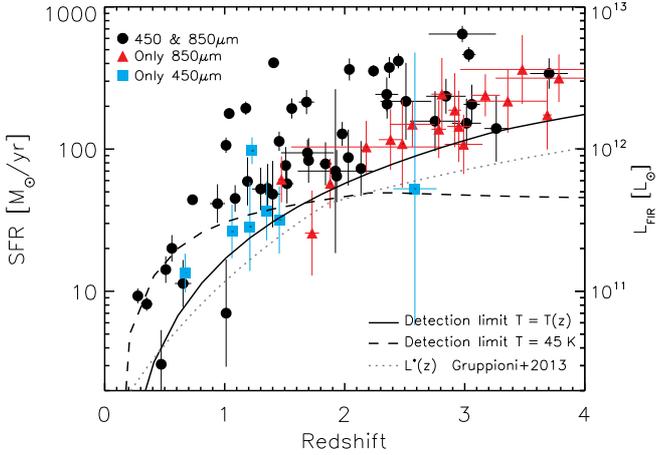}
  \caption{SFR ($\propto L_{\rm IR}$) as a function of redshift for galaxies with optical counterparts. The solid line represents the sensitivity
  limit of our survey   (considering both wavelengths)  assuming a modified blackbody with a dust temperature increasing with redshift 
  (see \S\ref{dust_temp_secc}), and the dashed line  a model with fixed dust temperature  at $T_{\rm d}=47\ K$ (the average temperature of our sample). As
  can be seen, the expected detection limit predicted by the SED with variable temperature is in better agreement with the measurements. 
  The depth of our survey allows us to detect galaxies with SFRs down to $\sim50\ \rm M_\odot\ {\rm yr}^{-1}$  at $z\approx1-2$ and we are sensitive to 
  $\rm SFR>150\ \rm M_\odot\ {\rm yr}^{-1}$   at any redshifts.
  The dotted line is $L^*(z)$ as reported by \citet{2013MNRAS.432...23G}, which is slightly below (less than a factor of $1.5$) our detection limit.}
  \label{z_SFR}
\end{figure}

\subsubsection{Dust temperature}\label{dust_temp_secc}

\begin{figure}
  \includegraphics[width=90mm]{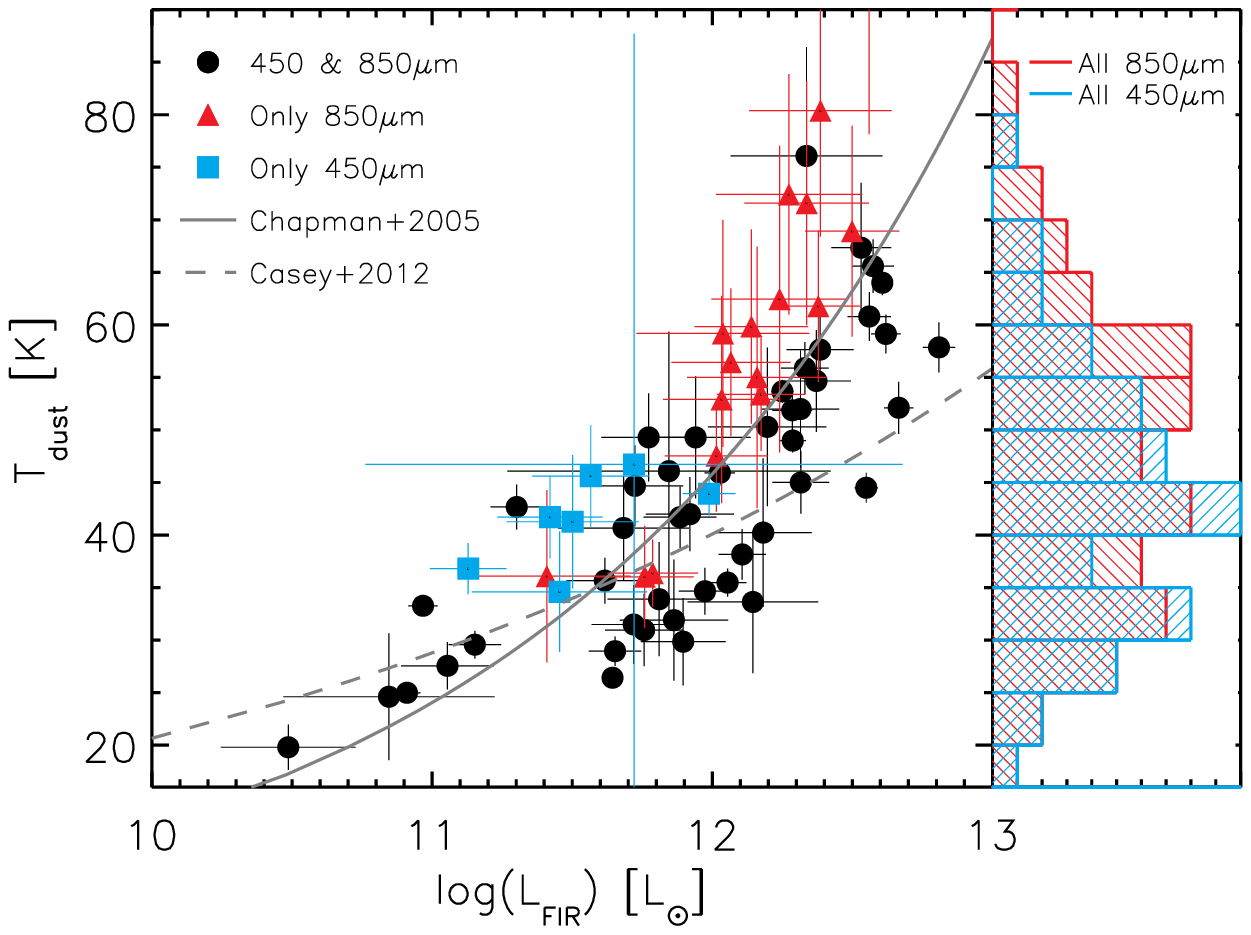}
  \caption{{\it Left:} Relation between dust temperature ($T_{\rm d}$) and IR luminosity ($L_{\rm IR}$), derived from a modified blackbody SED fitting at fixed redshift. 
  The solid line is the relation reported by \citet{2005ApJ...622..772C}, which qualitatively reproduces our measurements, and the dashed line represents 
  the relation found by  \citet{2012ApJ...761..139C}. {\it Right:} Dust temperature distributions for all the 450 and 850\! $\micron$-detected sources represented 
  by the blue and red histograms, respectively.  The mean dust temperature for each sub-sample is   $\langle T_{\rm d}\rangle=43\pm12\ \rm K$
and $49\pm15\ \rm K$ for the 450 and 850\! $\micron$ galaxies, respectively.  } 
  \label{temps_lum}
\end{figure}

The mean dust temperature for the whole sample is $\langle T_{\rm d}\rangle=47\pm15\ \rm K$, which is in agreement with previous studies. For example, \citet{2015ApJ...806..110D} found, using the MAGPHYS code (\citealt{2008MNRAS.388.1595D}), an average dust temperature of $T=43\pm2\ \rm K$ for 
the $870\ \micron$ ALMA-detected sources 
from the ALESS survey (\citealt{2013ApJ...768...91H}; \citealt{2013MNRAS.432....2K}), which cover a similar range
of redshifts and flux densities. On the other hand, \citet{2013MNRAS.436..430R} derived a mean temperature of $T_{\rm d}=42\pm11\ \rm K$ for 
$450\ \micron$-selected galaxies from the S2CLS in COSMOS and UDS, which have roughly the same depths as our maps.  These temperatures are, however, 
hotter than the ones derived by \citet{2014MNRAS.438.1267S} for the  ALESS sample when using modified blackbody fits, which have a median value 
of $\approx32\ \rm K$. These differences may be partially 
due to differences in the modified blackbody distributions, i.e. optically thin vs optically thick,  dissimilar emissivity indices, etc., as discussed
in \S\ref{dust_prop_secc}.

As shown in Fig. \ref{temps_lum}, the dust temperature follows the well-known temperature-luminosity relation found in previous 
studies (e.g. \citealt{2005ApJ...622..772C}; \citealt{2006ApJ...650..592K}; \citealt{2009MNRAS.393..653C}; \citealt{2012A&A...539A.155M};
\citealt{2013MNRAS.431.2317S}; \citealt{2014MNRAS.438.1267S}; \citealt{2016ApJ...833..103H}). Our results are consistent with the relation 
found by \citet{2005ApJ...622..772C} ($T_{\rm d}\propto L_{FIR}^{0.28}$; solid line in the figure). This slope is steeper than the one
reported by \citet{2012ApJ...761..139C} ($T_{\rm d}\propto L_{FIR}^{0.14}$; dashed line in Fig \ref{temps_lum}), however, their sample 
peaks at lower redshift ($\sim0.85$ vs $\sim2.3$) and it is selected at shorter wavelengths, so the differences can be explained by 
selection effects (see discussions by \citealt{2011MNRAS.411..505C}, \citealt{2011MNRAS.417.1192M}, and \citealt{2012ApJ...761..139C}).

Interestingly, if we inquire into the dust temperature of the 450 and  850\! $\micron$-detected galaxies independently, we find 
that both sub-samples show similar values. The dust temperature distributions for the 450 and 850 \! $\micron$-detected galaxies are 
represented in the right side of Fig. 
\ref{temps_lum}. These distributions are not statistically different from each other, with mean dust temperatures of $\langle T_{\rm d}\rangle=43\pm12\ \rm K$
and $49\pm15\ \rm K$ for the 450 and 850\! $\micron$ galaxies, respectively. This supports the idea that the redshift is the  main difference between both 
populations.

We also find an evolution of the dust temperature, with higher temperatures at higher redshifts, as is clearly seen in Fig. \ref{z_temps}.
This evolution has been reported in previous observational works (\citealt{2012ApJ...760....6M}; \citealt{2014A&A...561A..86M};
\citealt{2015ApJ...800...20G}; \citealt{2015ApJ...814....9K}; \citealt{2016ApJ...833..103H}), as well as in 
theoretical predictions (\citealt{2012MNRAS.426.2142L}; \citealt{2017MNRAS.tmp..172C}). The best-fitted linear model to our data is
described by $T_d=12(1+z)+11\rm\ K$ (see Fig. \ref{z_temps}), although with large scatter. As  a comparison, we also plot in the figure
the relations found by other authors
(\citealt{2013MNRAS.436.1919C}; \citealt{2015A&A...573A.113B}). With this evolution we can reproduce the luminosity detection limit seen in 
our data (see Fig. \ref{z_SFR}), and this also explains the systematic bias between our FIR photometric redshifts and the optical
redshifts $z_{\rm opt}$ (see \S\ref{z_dist_secc}). 

However, our luminosity detection limit is not flat at all redshifts. This selection effect 
along with the temperature-luminosity relation may produce an aparent redshift-temperature evolution in our sample, with a similar trend as the one in Fig. \ref{z_temps}, since the 
less luminous galaxies (which have colder dust temperatures) can only be detected at low redshift. To discard 
that this evolution is not produced only by selection effects, we perform the following simulation. First, we assume the dust temperature-luminosity relation of 
\citet{2005ApJ...622..772C}, which reproduces well our measurements (Fig. \ref{temps_lum}). Second, we estimate the flux density at 450 and 850\! $\micron$ for each 
simulated source based on the dust temperature, luminosity, and redshift, using equation \ref{eq:mbb}. The redshift of each source is 
chosen randomly (i.e. without any dependence on dust-temperature or luminosity) from a distribution that resembles those described in \S \ref{z_dist_secc}. Then, we impose our flux detection limit at each wavelength to consider only those sources that would be 
detected in our observations and, finally, we look for any relation between the dust temperature and redshift caused by selection effects. The results of these 
simulations are represented by the grey region in Fig. \ref{z_temps}. Indeed, the combination of our detection limit and the dust temperature-luminosity relation introduces
a trend between dust temperature and redshift, however, the relation found in the real data is much steeper (Fig. \ref{z_temps}) and cannot be explained only by selection effects. 

This implies that there is a more fundamental relation between $L_{\rm IR}$, $z$, and
$T_{\rm d}$ (and probably other parameters such as $M_{\rm dust}$ or sizes). This should be targeted in future studies of complete samples of
galaxies with spectroscopic redshifts in order to understand the physical processes behind this evolution. Some possible explanations include 
different metallicities at different redshifts (e.g. \citealt{2014A&A...561A..86M}),  evolution of the size of the star-forming region where 
galaxies at high redshifts are more compact and then hotter (e.g. \citealt{2016ApJ...833..103H}; \citealt{2018NatAs...2...56Z}), or a combination
of different effects (e.g. \citealt{2017MNRAS.tmp..172C}).

\begin{figure}
  \includegraphics[width=90mm]{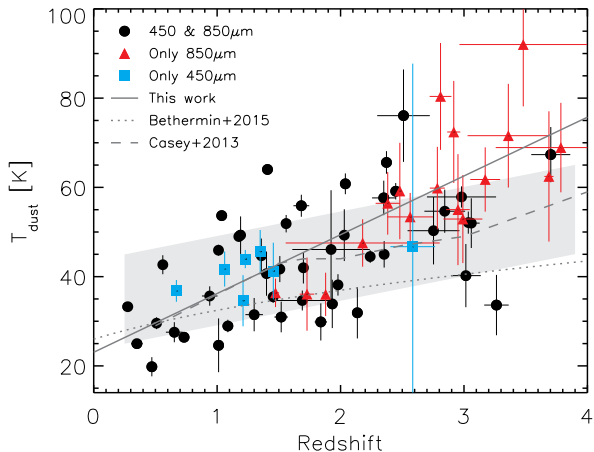}
  \caption{Dust temperature against redshift for sources with optical counterparts. The dust temperature has been estimated through a greybody SED fitting
  at fixed redshift with fixed $\beta=1.6$. The dashed line represents the  best linear fit for our sample, with dust temperature increasing with 
  redshift (see discussion in \S\ref{dust_temp_secc}).
  As it can be seen, the galaxies detected at just one single band (blue and red circles for 450 and 850 \! $\micron$-only detected sources, respectively) 
  follow the same trend that those detected at both wavelengths (black circles).  For comparison, we also plot the relation found by \citet{2013MNRAS.436.1919C} 
  and \citet{2015A&A...573A.113B}. The grey shaded region represents the aparent evolution of dust temperature with redshift 
  originated by selection effects, specifically, due to the combination of our luminosity detection limit and the dust temperature-luminosity relation (see main text
  for details of the simulation). However, this effect cannot totally explain  the correlation found in the data, which implies a real evolution of dust 
  temperature with redshift. }
  \label{z_temps}
\end{figure}

\subsection{Stellar masses \& the main-sequence}\label{main_seq_secc}

\begin{figure*}
  \includegraphics[width=170mm]{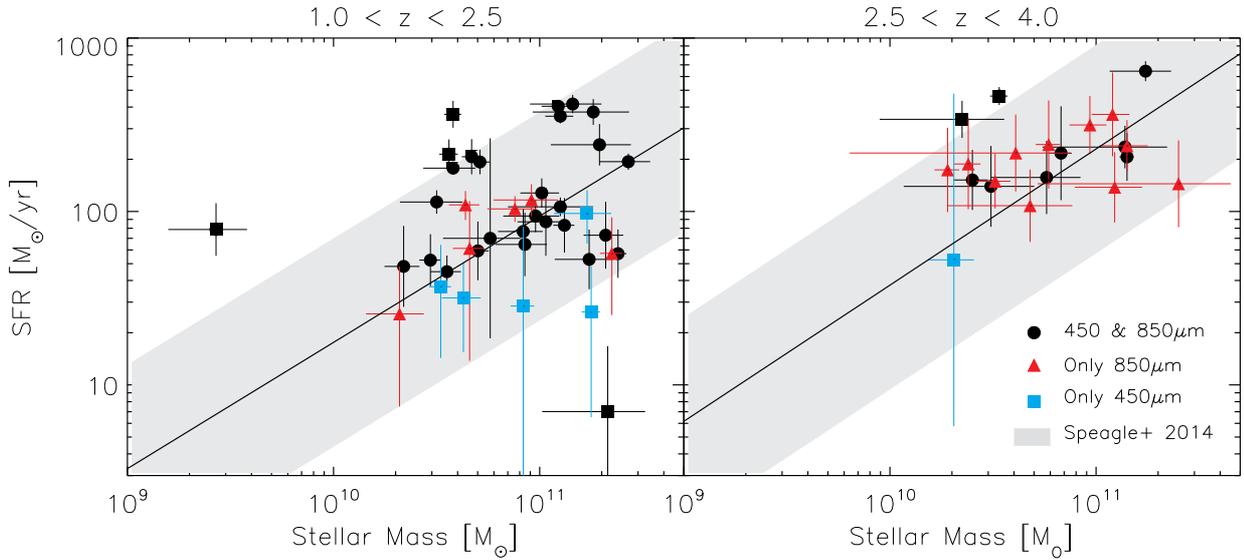}
  \caption{The SFR as a function of stellar mass for those galaxies with optical counterparts in two different redshift bins indicated at the top of the of each panel.
  The solid line represents the main-sequence of star-forming galaxies at $z\sim1.5$ ({\it left}) and $z\sim3$ ({\it right}) derived from a compilation of different studies
  by \citet{2014ApJS..214...15S} and  the shaded area shows the $\pm3\sigma$ scatter over the defined average. As it can be seen,
  most SMGs lie within  a factor of $\pm3\sigma$ of the main-sequence, represented   by the grey shaded region.}
  \label{main_seq}
\end{figure*}

In this section we explore the connection between star formation rate and stellar mass for our galaxies. We adopt the stellar masses from the 3D-HST catalogue 
(\citealt{2014ApJS..214...24S}; \citealt{2016ApJS..225...27M}), which are derived from SED fitting with 
{\sc FAST} (\citealt{2009ApJ...700..221K}) using the {\sc `Z-BEST'} redshifts (see \S\ref{z_dist_secc}). The SEDs are based on \citet{2003MNRAS.344.1000B} stellar
population synthesis (SPS) models with a \citet{2003PASP..115..763C} IMF. For the small fraction of sources which are outside of the 3D-HST
coverage (13 out of 71), we use the stellar mass reported by \citet{2011ApJS..193...30B} derived using the same SPS models and the same IMF.  
A comparison between the stellar mass derived for the SCUBA-2 common sources in these catalogues shows that both estimations are in reasonable agreement with an average
$\log(M_{\rm *3DHST}/M_{\rm *Barro2011})\approx0.2$ (without any systematic offset). 

The mean stellar mass for the galaxies with optical counterpart is $9\pm0.6\times10^{10}\ M_\odot$, and $>90$ per cent of these galaxies have $M_*\ga1\times10^{10}\ M_\odot$. 
This is in good agreement with the estimations of the ALESS sources by \citet{2015ApJ...806..110D}, with a median stellar mass of $8.9\pm0.1\times10^{10}\ M_\odot$,
as well as with other studies of galaxies with similar flux densities (e.g. \citealt{2013ApJ...768..164A}; \citealt{2014ApJ...788..125S}; \citealt{2016MNRAS.458.4321K};
\citealt{2017MNRAS.469..492M}). This is also in agreement with the values found by \citet{2017MNRAS.466..861D} for the sources detected in the ALMA image of 
the Hubble Ultra Deep Field, where $\sim80$ per cent of the sources have $M_*\ga1\times10^{10}\ M_\odot$, even though the ALMA map is deeper and was done at
1.3\! mm. 

In Fig. \ref{main_seq} we plot the SFR as a function of stellar mass for sources in two redshift bins ($1.0\le z<2.5$ and $2.5\le z<4.0$). 
This shows the place occupied by our galaxies relative to the `main-sequence' of star-forming galaxies, for which we adopt the parameterization reported by 
\citet{2014ApJS..214...15S} based on a compilation of several studies from the literature covering a wide range of stellar masses up to $z\sim6$. 
In the low-redshift bin ($1.0\le z<2.5$), we are sensitive down to SFR$\sim50\ M_\odot\ {\rm yr}^{-1}$ and $M_*\sim3\times10^{10}\ M_\odot$, and in the 
high-redshift bin down to SFR$\sim150\ M_\odot\ {\rm yr}^{-1}$. This allows us to sample the typical parameter space of the `main-sequence'  
of star-forming galaxies (for these redshifts and stellar masses) and hence to study the nature of our galaxies in this context.  
It can be seen that at these redshifts, most of our galaxies ($\sim85$ per cent) lie on the high-mass end of the `main-sequence' within a factor of $\pm3\sigma$, 
where we assume as $1\sigma$ the 0.2\ dex intrinsic scatter reported by \citet{2014ApJS..214...15S}. This result is  consistent with previous studies 
(e.g. \citealt{2012A&A...541A..85M}; \citealt{2016MNRAS.458.4321K}; \citealt{2017MNRAS.469..492M}; \citealt{2017MNRAS.466..861D}; \citealt{2017A&A...599A.134S}), 
and it has been predicted by 
some cosmological simulations (e.g. \citealt{2010MNRAS.404.1355D}; \citealt{2015Natur.525..496N}).  On the other hand, we can also represent the
`main-sequence' of star-forming galaxies in terms of the specific SFR (sSFR=SFR$/M_*$), and take into account its evolution with redshift. This
is shown in Fig. \ref{main_seq2}, from which it is clearly seen that our galaxies (even those at $z<1$) lie on this relation and follow the same redshift evolution. 

% 12/95 galaxies with X-ray detection - AGN fraction of \sim 10%?

Only a small fraction ($\la15$ per cent) of our faint SMGs are above the main-sequence. Actually, from the 8 sources above it three have at least another optical
galaxy within our search radius, and therefore, their exceeded SFR may be explained by the blending of multiple sources. 
This fraction would be even lower if we adopt the 0.3\ dex observed scatter of the `main-sequence' instead of the intrinsic one (\citealt{2014ApJS..214...15S}).  This suggests that most SMGs can be fully explained as the most massive star-forming main-sequence galaxies, as also discussed by \citealt{2017MNRAS.469..492M}.

% Interestingly, thanks to the depth of our survey, we are also able to detect some galaxies which lie below the `main-sequence'.  This  indicates 
% that future deeper (sub-)mm surveys will be able to detect complete samples of sub-$L^*$ galaxies at high-redshifts.\\

\begin{figure}
  \includegraphics[width=90mm]{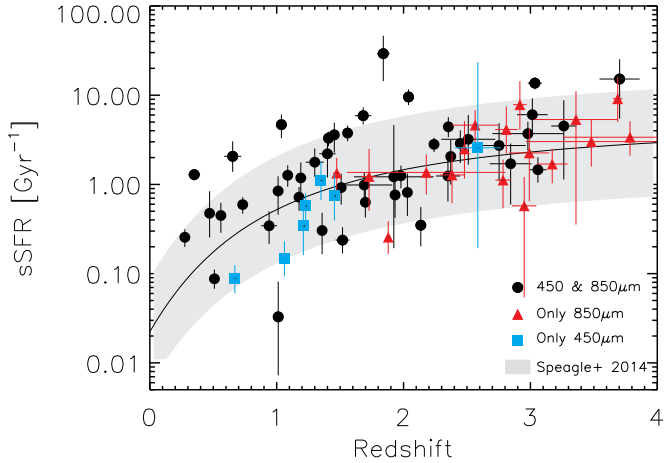}
  \caption{sSFR as a function of redshifts for all the galaxies with optical counterparts.  The solid line represents the main-sequence of star-forming galaxies 
  for $M_*=9\times10^{10}$\! M$_{\odot}$ (\citealt{2014ApJS..214...15S}). 
  Most of our galaxies ($\sim85$ per cent) lie within a factor of $\pm3\sigma$ of the  main-sequence  (indicated by the grey shaded region). }
  \label{main_seq2}
\end{figure}

\subsection{The IRX-$\bmath{\beta}$ relation}

The Infrared excess (${\rm IRX}=\log[L_{\rm IR}/L_{\rm UV}]$) to UV-slope ($\beta$, $f_\lambda\propto\lambda^\beta$) relation, links the amount of dust absorption
measured in the UV to the amount of infrared re-emission. It has been shown that local starburst galaxies follow a tight  IRX-$\beta$ relation (e.g. 
\citealt{1999ApJ...521...64M}; \citealt{2000ApJ...533..682C}), which may be used to infer dust obscuration, and hence total SFRs, when IR data is not available. 
However, this relation is uncertain at high redshifts, with some studies reporting galaxies consistent with the local relation 
(e.g. \citealt{2012ApJ...744..154R}, \citealt{2014ApJ...792..139T}; \citealt{2017MNRAS.467.1360B}) while others 
showing sources above (e.g. \citealt{2012ApJ...759...28P}, \citealt{2013A&A...554L...3O}) and below 
(e.g. \citealt{2015Natur.522..455C}; \citealt{2016A&A...587A.122A}; \citealt{2016ApJ...833...72B}; \citealt{2017arXiv170304535P}) of this relationship. These high-redshift studies have been done with 
samples selected with different criteria (e.g. UV-selected vs IR-selected galaxies), which could introduce some extra effects (\citealt{2014ApJ...796...95C}). 
Here we use our deep survey to constrain the dust absorption properties of faint SMGs in the context of the IRX-$\beta$ 
relation. For this analysis we limit the sample to those sources detected within the 3D-HST catalogue, which provides 
estimations for the UV spectral slope. This quantity is determined from a power-law fit of the form 
$f_\lambda\propto\lambda^\beta$ (\citealt{2014ApJS..214...24S}). The IRX is determined as the ratio between the SFR 
determined from the IR data (see \S\ref{lum_secc}) and the SFR derived from the rest-frame $1600\ \AA$ luminosity 
(not corrected for dust attenuation, \citealt{2014ApJS..214...24S}). 

Fig. \ref{irx_beta} shows the locus occupied by 
our galaxies in the IRX-$\beta$ plane, colour-coded according to their IR luminosities.  For comparison, we also plot
the \citet{1999ApJ...521...64M} relation usually adopted  for starburst galaxies (see also \citealt{2000ApJ...533..682C}) 
and its revised version corrected for aperture effects (\citealt{2012ApJ...755..144T}), along with the relation found 
in the Small Magallanic Cloud (SMC; \citealt{1998ApJ...508..539P}).  This figure suggests that there is a variation 
in the IRX-$\beta$ relation as a function of IR luminosity ($\propto{\rm SFR}$), where galaxies with 
$L_{\rm IR}\ga1\times10^{12}\ L_\odot$ (the most luminous objects) lie above of the corrected starburst relation, 
while source with $L_{\rm IR}\la1\times10^{12}\ L_\odot$ preferentially lie below it, in better agreement with the SMC 
relation (although with a large scatter).  In Figure \ref{irx_beta}, we also plot as
comparison other IR-detected galaxies with IR luminosity similar to our less luminous sources (\citealt{2015Natur.522..455C};
\citealt{2017arXiv170304535P}), which show similar results.
The same trend as a function of IR luminosity was reported by 
\citet{2014PhR...541...45C} with a break luminosity of $\approx10^{11.5}$, and by \citet{2015ApJ...806..110D} who show 
tentative evidence of lower dust attenuation for the lowest luminous galaxies.  Exploring other dependencies, we 
also find a variation as a function of dust temperature and sSFR, which is not surprising given the correlation between these
parameters and the IR luminosity.

The explanation for this deviation from the nominal relation is not entirely clear. The interpretions of different 
observational results and some theoretical models span different scenarios. For example, a more abundant population of 
young O and B stars for the bluer galaxies (\citealt{2014ApJ...796...95C}), dust composition and enrichment
(\citealt{2016MNRAS.462.3130M}; \citealt{2016arXiv160407402S}), geometry effects (\citealt{2015ApJ...806..110D}), different
star formation histories (\citealt{2004MNRAS.349..769K}), or a combination of these factors (\citealt{2016ApJ...827...20S}). 
Interestingly, some of our galaxies lie below the SMC relation, even when this curve is generally thought to be a limiting
case for star-forming galaxies (e.g. \citealt{1998ApJ...508..539P}). This could be evidence of older stellar populations in these
objects (e.g. \citealt{2004MNRAS.349..769K,2017MNRAS.472.2315P}), however, this scenario is not suitable for the 
$z\sim5-6$ galaxies reported by \citet{2015Natur.522..455C} which are expected to be dominated by very young systems.
In any case, this implies a non-universal IRX-$\beta$ relation.

\begin{figure}
  \includegraphics[width=90mm]{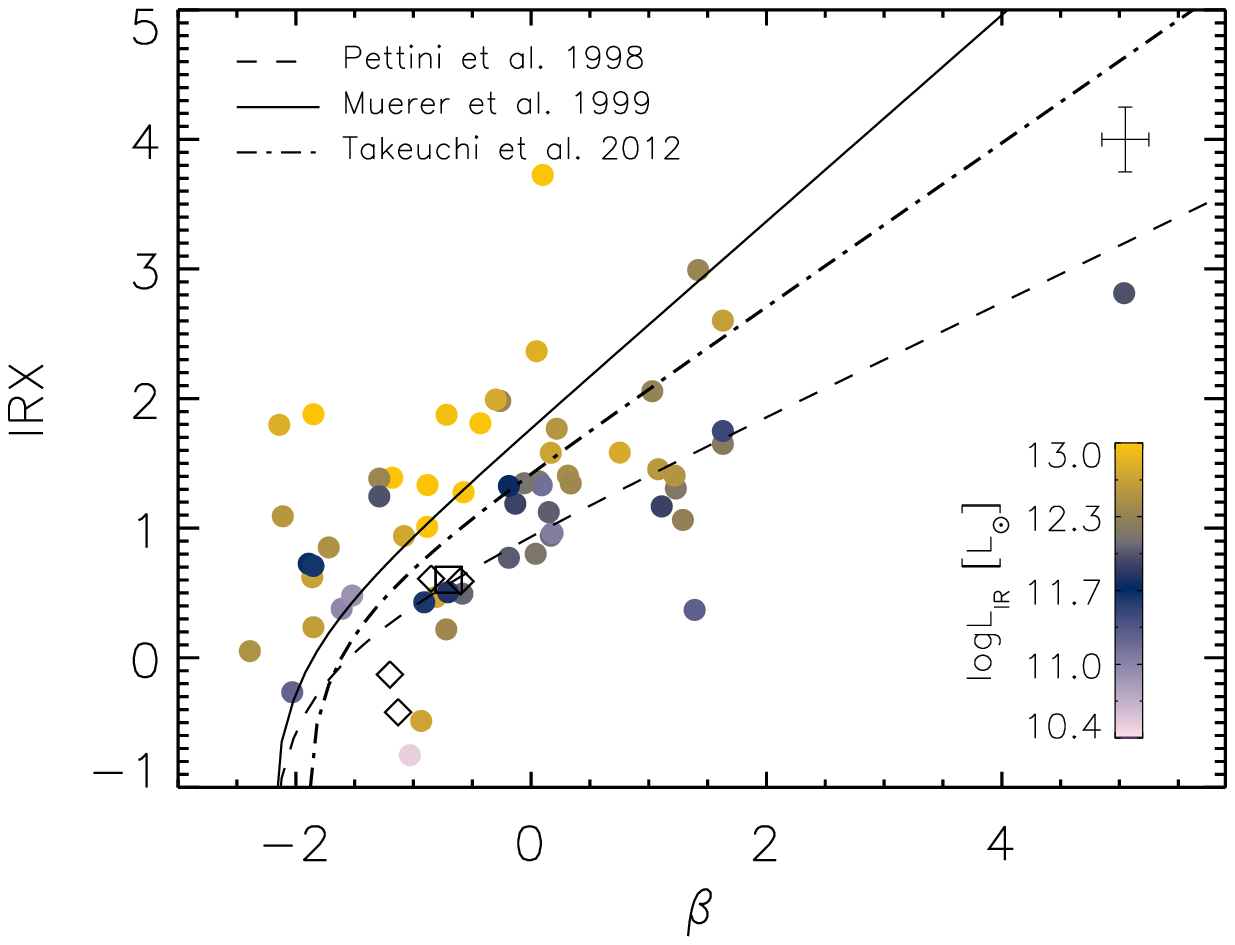}
  \caption{Infrared excess (IRX) versus the UV-continuum slope ($\beta$) for galaxies with optical counterpart. The points are 
  colour-coded by their   infrared luminosity. The solid line shows the original IRX-$\beta$ relation for starburst galaxies 
  (\citealt{1999ApJ...521...64M}), which was then   aperture-corrected by \citet{2012ApJ...755..144T} (dot-dashed line), while 
  the dashed line represents   the SMC extinction curve (\citealt{1998ApJ...508..539P}). The typical uncertainty of our values 
  is represented on the top right-hand corner. The small diamonds represent ALMA observations of $z\sim5$   galaxies with similar 
  luminosities ($L_{\rm FIR}\approx 3\times10^{11}\rm\ L_\odot$, \citealt{2015Natur.522..455C}),   and the square is  the 
  measurement of a lensed galaxy at $z\approx4.1$ with $L_{\rm FIR}\approx1\times10^{11}\rm\ L_\odot$ detected by the LMT 
  (\citealt{2017arXiv170304535P}).}
  \label{irx_beta}
\end{figure}

\subsection{Morphologies}

We use the morphological classification derived by \citet{2012ApJS..203...24V} and \citet{2015ApJS..221....8H} in order to study the morphological
properties of our sample. These studies consist of two-dimensional axisymmetric S\'ersic models fitted with {\sc GALFIT} (\citealt{2012ApJS..203...24V}), and
visual-like classifications estimated using Convolutional Neural Networks (\citealt{2015ApJS..221....8H}). These methods have been applied to the 
5 CANDELS fields (\citealt{2011ApJS..197...35G}) using the HST {\it H$_{\rm 160}$}-band, which probes the optical rest-frame at the typical
redshifts of our sources, and therefore, the mass-dominant component, rather than the high surface-brightness features which commonly dominate in the UV. 

\begin{figure*}
\includegraphics[width=180mm]{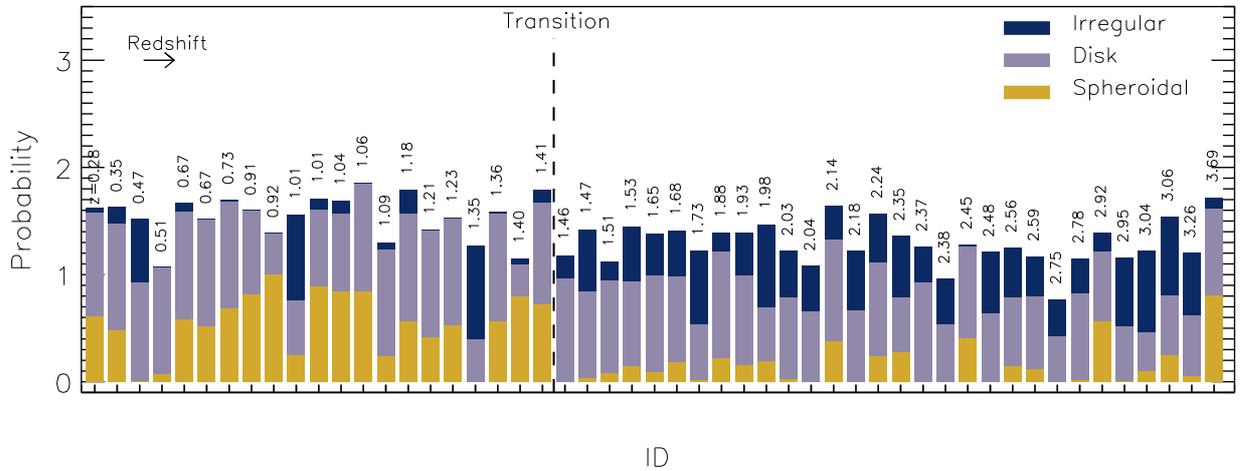}
\caption{Probability of flagging a galaxy as spheroidal ({\it gold}),  disk ({\it purple}), or  irregular ({\it blue}), based on a visual-like classification 
in the {\it H$_{\rm 160}$}-band estimated through a neural network algorithm (\citealt{2015ApJS..221....8H}). The probabilities may add up to more than 1 because 
the categories are not mutually exclusive. The redshift of each galaxy is indicated in the top of each bar 
and galaxies are sorted from low to high redshift. Most of the sources have a higher probability of being classified as disks. However, the higher redshift sources
($z\ga1.4$) show also a significant probability of having some irregularities, while the lowest redshift galaxies ($z\la1.4$) have also a significant probability 
of having a spheroidal component. This may be interpreted as evidence of  structural evolution of SMGs and supports the idea that these objects are the progenitors of
 massive elliptical galaxies.}
\label{visual_class}
\end{figure*}

The visual-like classification of \citet{2015ApJS..221....8H}, which is based on neural networks trained to reproduce the visual morphologies published 
by \citet{2015ApJS..221...11K}, are able to predict the results of experts
classifiers with a bias close to zero and a $\sim10$ per cent scatter, with a misclassification fraction less than 1 per cent. Each galaxy has five associated real 
numbers  which correspond to the frequencies at which expert classifiers would have flagged the galaxy as having a spheroid, having a disk, presenting an irregularity,
being compact or a point source, and being unclassifiable. Based on these numbers, we show in Fig. \ref{visual_class}  the probability of our galaxies to be classified 
as spheroidal, disk, or irregular (the probability of being classified as point source or unclassifiable is negligible for most of our sources). As it can be seen, 
most of our galaxies have a larger probability of being classified as a disk at all redshifts, but we note an interesting transition at $z\sim1.4$ with galaxies at
high-redshifts showing also a significant probability of being classified as irregulars, or as spheroidals at lower redshifts.

Following the classification example of \citet{2015ApJ...809...95H} we defined the following morphological classes based on the aforementioned probabilities:\\

\noindent$\bullet$ Pure disks: $f_{\rm disk}>2/3$ AND $f_{\rm sph}<2/3$ AND $f_{\rm irr}<2/3$\\
$\bullet$ Pure bulges: $f_{\rm disk}<2/3$ AND $f_{\rm sph}>2/3$ AND $f_{\rm irr}<2/3$\\
$\bullet$ Irregular disks: $f_{\rm disk}>2/3$ AND $f_{\rm sph}<2/3$ AND $f_{\rm irr}>2/3$\\
$\bullet$ Disk$+$bulges: $f_{\rm disk}>2/3$ AND $f_{\rm sph}>2/3$ AND $f_{\rm irr}<2/3$\\
$\bullet$ Irregulars/mergers: $f_{\rm disk}<2/3$ AND $f_{\rm sph}<2/3$ AND $f_{\rm irr}>2/3$\\

\noindent where $f_{\rm disk}$, $f_{\rm sph}$, $f_{\rm irr}$, are the probabilities of being classified as a disk, spheroidal, or irregular, respectively. Based on
this scheme, we can understand the transition in Fig. \ref{visual_class} as a high fraction of irregular disks at high redshifts and a high fraction of disks$+$bulges 
at low redshift. This is clearly seen in Fig.
\ref{morph_vs_z}, where we plot the fraction of pure disks, irregular disks, and disk$+$bulges, as a function of redshift (we have not found pure bulges or 
irregular/mergers in our sample). From $z\approx3$ to 0.5  the galaxies move from being dominated by irregular disks to being dominated by disks$+$bulges. 

 If these galaxies follow the same evolutionary path, this implies a morphological transformation from high to low redshift which results in the growth of 
a bulge component. This results supports, morphologically, the scenario of SMGs as progenitors of present-day massive elliptical galaxies. This  has been inferred 
before by previous studies based on  indirect evidence as the number space density, the star formation history and stellar masses, clustering, etc. 
(e.g. \citealt{1999ApJ...518..641L};  \citealt{2002MNRAS.331..495S}; \citealt{2004MNRAS.355..424T}; \citealt{2006MNRAS.371..465S}; \citealt{2012MNRAS.421..284H};
\citealt{2014ApJ...788..125S}).  Similar morphological evolution has also been reported in the studies of optically detected massive galaxies 
($M_*\ga1\times10^{11}$\! M$_{\odot}$, e.g. \citealt{2013MNRAS.433.1185M}; \citealt{2014MNRAS.444.1001B}; \citealt{2015ApJ...809...95H}), which supports the scenario
that SMGs can be explained as the most massive star-forming main-sequence galaxies (e.g.  \citealt{2017MNRAS.469..492M}).

One important caveat is that surface brightness and other redshift issues, such as the sampling of different rest-frame wavelengths or the evolution 
of the galaxies' apparent sizes, may affect the morphological classification, particularly the detection of bulges at high redshifts (which could lead us 
to a similar picture to the one presented in Fig. \ref{morph_vs_z}). However, \citet{2015ApJS..221....8H} have shown that their method is able to reproduce
the visual classification with a bias close to zero up to (at least) $z\sim2.7$. Furthermore, the transition occurs abruptly between $z\approx1-2$, 
where the H$_{\rm 160}$-band is still sensitive to the optical rest-frame from older stars.

\begin{figure}
  \includegraphics[width=90mm]{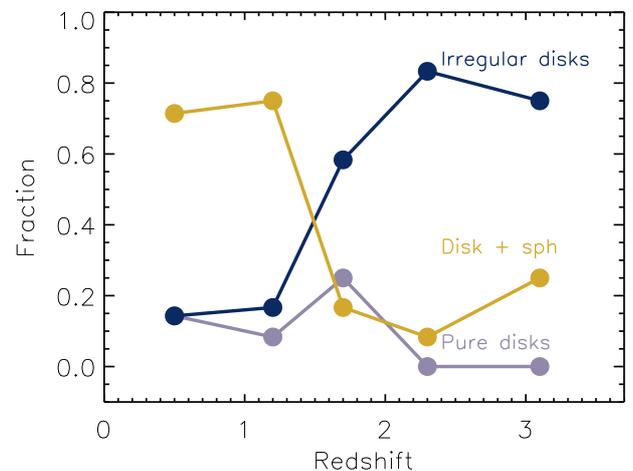}
  \caption{Fraction of galaxies classified as `pure disks', `irregular disks' and  `disks + spheroids' (see definitions in the text) based on the visual-like 
  {\it H$_{\rm 160}$}-band classification of \citet{2015ApJS..221....8H}. Although  most of the galaxies show a higher probability of being classified as disks, 
  there is a transition from   `irregular disks' to disks with a spheroidal component at $z\sim1.4$.}
  \label{morph_vs_z}
\end{figure}

Similar results can be obtained using the parameters of the best-fitting S\'ersic models. Structural properties (i.e. effective radii and S\'ersic indices; \citealt{1963BAAA....6...41S}) are taken
from the public catalogue released by \citet{2012ApJS..203...24V}, which were obtained with {\sc GALFIT} and {\it H$_{\rm 160}$}-band images. 
The median S\'ersic index of our sample is $n=1.4^{+0.3}_{-0.1}$ (see S\'ersic index distribution in Fig. \ref{sersic_index}) with a median half-light radius of $r_{1/2}=4.8\pm0.4$\! kpc, where the errors have been estimated via bootstrap resampling. 
This is in good agreement with the results found by
\citet{2013MNRAS.432.2012T} and \citet{2015ApJ...799..194C}, who also studied the morphology of SMGs in NIR bands. They reported median values of $n=1.0\pm0.1$ and 
$r_{1/2}=4.1\pm0.5$\! kpc, and $n=1.2\pm0.3$ and $r_{1/2}=4.4^{+1.1}_{-0.5}$\! kpc, respectively. These values support the conclusion derived from the visual-like classification, 
that the dominant component in our galaxies is disc-like, as discussed by \citet{2013MNRAS.432.2012T}. 

In order to study the morphological transition described above
(Fig. \ref{morph_vs_z}), we estimate the median S\'ersic index in two redshift bins. The galaxies at higher redshifts ($z>1.4$) have a median value of $n\approx1$, 
while the low-redshift galaxies ($z\le1.4$) show a larger value of $n\approx2$. This increases in the median S\'ersic index from high to low redshift also suggests a 
morphological transformation, which implies the formation of a bulge component. We emphasize that this conclusion is supported by two completely independent 
morphological classification methods and represents morphological evidence for an evolutionary link in the direction of SMGs evolving towards local elliptical galaxies. 
The mechanism of bulge formation, however, 
cannot be deduced from our data, since we cannot discriminate between gas rich major mergers and violent disk instabilities, or other scenarios like minor mergers.

% . We use only a
% classification in the H band (F160W) since the differences in
% the derived (broad) morphologies when using other filters are
% very small as shown in Kartaltepe et al. (2014).

% Huertas-company 2016
% Strong dissipative processes
% such as very-gas rich mergers or violent disk instabilities
% are known to rapidly bring a large amount of gas into the
% central parts of the galaxy, leading to a massive, compact
% and dense remnant as observed. Alternatively, they might
% be the result of quenching of small star-forming systems
% at higher redshifts

% by exploring the morphologies of galaxies in the optical we are
% not biased towards very blue features, a problem which could
% be part of the explanation for the differences in studies so
% % far.

% Mortlock 2013
% Furthermore, the H160
% imaging probes optical light in the redshift range z = 1 − 3,
% making this data ideal for studying the visual morphologies
% of high redshift galaxies, unlike several past surveys which
% have imaged galaxies in the rest-frame UV. Rest-frame UV
% light is dominated by star formation features which may not
% represent well the underlying light distributions.

\begin{figure}
  \includegraphics[width=90mm]{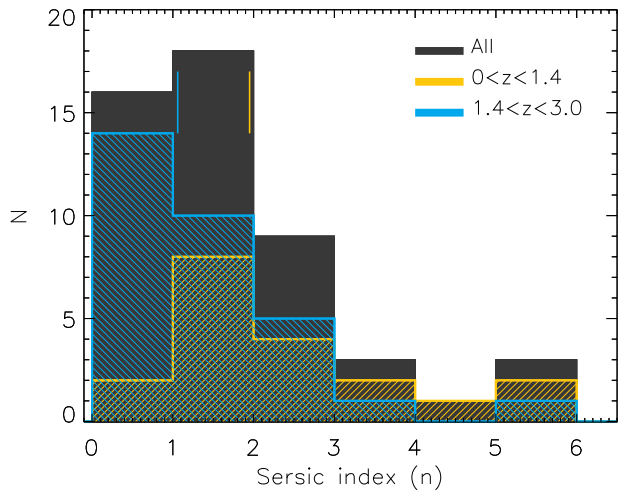}
  \caption{The S\'ersic index distribution for our sources with optical counterparts derived from a two-dimensional modelling of the HST {\it H$_{\rm 160}$}-band images 
  (\citealt{2012ApJS..203...24V}). The sources have been divided into two sub-samples: galaxies with $z\le1.4$ are represented by the blue histogram and
  the sources with $1.4<z<3$ by the yellow one. The vertical lines represent the median value for each sub-sample, where the lowest redshift galaxies show
  larger sersic index.}
  \label{sersic_index}
\end{figure}

\section{Conclusions}

We have used the rich multi-wavelength dataset available in the EGS field to study the physical properties of faint submillimeter galaxies detected 
in deep SCUBA-2 observations at 450 and 850\ $\micron$ (\citealt{2017MNRAS.464.3369Z}). This survey, together with other similar deep SCUBA-2 maps, 
represents one of the deepest blank-field observations achieved with a single-dish telescope, which allows us to detect galaxies with $\rm SFR\ge150\ M_\odot\ {\rm yr}^{-1}$
at all redshifts.  Our sample consists of both, 450 and 850\! $\micron$-selected galaxies,
which increases the completeness and minimizes the selection effects associated with single-band observations (as discussed by \citealt{2013MNRAS.436.1919C}). 
Additionally, this allows us to explore the physical differences between these populations. The final sample 
includes 95 sources: 56 (59\! \%) are detected at both wavelengths, 31 (33\! \%) are detected only at 850\! $\micron$, and 8 (8\! \%) are detected only 
at 450\! $\micron$.

Using radio and IR observations as an intermediate step to associate our SMGs with optical counterparts, we achieved a robust association rate of 75\! \%.
Combining optical photometric and spectroscopic redshifts (\citealt{2011ApJS..193...30B}; \citealt{2014ApJS..214...24S}; \citealt{2016ApJS..225...27M})
with FIR photometric redshifts for those galaxies with no counterpart (see \S\ref{z_dist_secc}) we derived the redshift distributions for both 450 and  
850\ $\micron$-selected galaxies, which have a median redshift of $1.66\pm0.18$ and $2.30\pm0.20$, respectively. This supports the scenario in which 
galaxies selected at different wavelengths
have different redshift distributions (see also \citealt{2014MNRAS.443.2384Z}; \citealt{2015A&A...576L...9B}).  Actually, all (except one) 
of the galaxies detected only at 450\! $\micron$ (i.e. with no 850\! $\micron$ detection) lie at $z\la1.5$, while all the galaxies detected only at 850\! $\micron$ lie at 
$z\ga1.5$.

For those galaxies with optical counterparts we derived infrared luminosities, star formation rates, and dust temperatures. These, combined with their stellar masses, allow
us to study the locus occupied by our galaxies relative to the `main-sequence' of star forming galaxies. Additionally, we also studied the dust absorption properties 
of these galaxies (in the context of the IRX-$\beta$ relation) as well as the optical rest-frame morphology. The main conclusions from these analyses are:

$\bullet$ The mean IR luminosity of our sample is $1.5\pm0.2\times10^{12}\ L_\odot$, which corresponds to a mean SFR of $150\pm20\ M_\odot\ {\rm yr}^{-1}$, covering 
a range from $L_{\rm IR}=0.3\times10^{11}-7.4\times10^{12}\ L_\odot$. 

$\bullet$ The mean dust temperature for the whole sample is $\langle T_{\rm d}\rangle=47\pm15\ \rm K$, and temperature monotonically increases
towards high redshift (see \S\ref{dust_temp_secc}). This $T_d-z$ relation is supported by: (1) higher dust temperatures at higher redshift 
(Fig. \ref{z_temps}); (2) the luminosity detection limit seen in our data (see Fig. \ref{z_SFR}) which is well fitted when we consider this evolution, and; (3)  
a trend where the FIR photometric redshifts are systematically lower than the optical ones at $z_{\rm opt}>2$ if a single dust temperature SED is assumed,
which can be translated into a difference between the real dust temperature of the source and that of the SED (as reported before by \citealt{2016ApJ...832...78I}).

$\bullet$ We found that most of our galaxies ($\sim85$ per cent) lie on the high-mass end of the `main-sequence' within a factor of $\pm3\sigma$, with a mean stellar mass of 
$9\pm0.6\times10^{10}\ M_\odot$.  

$\bullet$ In the context of dust absorption, we found a variation in the IRX-$\beta$ relation as a function of IR luminosity, where galaxies with 
$L_{\rm IR}\ga1\times10^{12}\ L_\odot$ lie on (or slightly above for the most luminous objects) the \citet{1999ApJ...521...64M} relation,
while sources with $L_{\rm IR}\la1\times10^{12}\ L_\odot$ are in better agreement with the SMC relation. A similar behaviour can be found as a function of
dust temperature and sSFR.

$\bullet$ Based on visual-like classifications of the HST {\it H$_{\rm 160}$}-band (\citealt{2015ApJS..221....8H}) we concluded that the dominant component for most 
of our galaxies is a disk-like structure, although we note an interesting transition at $z\sim1.4$. Galaxies move  from being irregular disks-dominated
at high redshifts to disks+bulges at 
low redshift, which morphologically supports the scenario of SMGs being the  progenitors of present-day massive elliptical galaxies. Similar conclusion were obtained
using two-dimensional axisymmetric S\'ersic models (\citealt{2012ApJS..203...24V}), where the highest redshift galaxies have median S\'ersic index of $n\approx1$ while the lowest redshift galaxies
show a larger value of $n\approx2$.

\section*{Acknowledgements}
{\small
We would like to thank the anonymous referee for a detailed report that has increased the clarity of the paper.
In addition, we thank to James Simpson for his suggestions along this work.
This research has been supported by Mexican CONACyT research grant
CB-2011-01-167291. JAZ acknowledges support from a CONACyT 
studentship and from the University of Texas at Austin College of Natural Sciences, and the hospitality at the University of Edinburgh.
The James Clerk Maxwell Telescope has historically been operated by the Joint Astronomy Centre on behalf of the Science
and Technology Facilities Council of the United Kingdom, the National Research Council of Canada, and
the Netherlands Organisation for Science Research. Additional funds for the 
construction of SCUBA-2 were provided by the Canada Foundation for Innovation. This work is based in part
on observations made with the {\it Spitzer Space Telescope}, which is operated by the Jet Propulsion Laboratory, 
California Institute of Technology under a contract with NASA. This research has made use of data from HerMES
project (http://hermes.sussex.ac.uk/). HerMES is a Herschel 
Key Programme utilising Guaranteed Time from the
SPIRE instrument team, ESAC scientists and a mission scientist.
}

%%%%%%%%%%%%%%%%%%%%%%%%%%%%%%%%%%%%%%%%%%%%%%%%%%



\bibliographystyle{mnras}
\bibliography{biblio} % if your bibtex file is called example.bib

\begin{thebibliography}{}
\makeatletter
\relax
\def\mn@urlcharsother{\let\do\@makeother \do\$\do\&\do\#\do\^\do\_\do\%\do\~}
\def\mn@doi{\begingroup\mn@urlcharsother \@ifnextchar [ {\mn@doi@}
  {\mn@doi@[]}}
\def\mn@doi@[#1]#2{\def\@tempa{#1}\ifx\@tempa\@empty \href
  {http://dx.doi.org/#2} {doi:#2}\else \href {http://dx.doi.org/#2} {#1}\fi
  \endgroup}
\def\mn@eprint#1#2{\mn@eprint@#1:#2::\@nil}
\def\mn@eprint@arXiv#1{\href {http://arxiv.org/abs/#1} {{\tt arXiv:#1}}}
\def\mn@eprint@dblp#1{\href {http://dblp.uni-trier.de/rec/bibtex/#1.xml}
  {dblp:#1}}
\def\mn@eprint@#1:#2:#3:#4\@nil{\def\@tempa {#1}\def\@tempb {#2}\def\@tempc
  {#3}\ifx \@tempc \@empty \let \@tempc \@tempb \let \@tempb \@tempa \fi \ifx
  \@tempb \@empty \def\@tempb {arXiv}\fi \@ifundefined
  {mn@eprint@\@tempb}{\@tempb:\@tempc}{\expandafter \expandafter \csname
  mn@eprint@\@tempb\endcsname \expandafter{\@tempc}}}

\bibitem[\protect\citeauthoryear{{Aguirre}, {Baker}, {Menanteau}, {Lutz}  \&
  {Tacconi}}{{Aguirre} et~al.}{2013}]{2013ApJ...768..164A}
{Aguirre} P.,  {Baker} A.~J.,  {Menanteau} F.,  {Lutz} D.,   {Tacconi} L.~J.,
  2013, \mn@doi [\apj] {10.1088/0004-637X/768/2/164}, \href
  {http://adsabs.harvard.edu/abs/2013ApJ...768..164A} {768, 164}

\bibitem[\protect\citeauthoryear{{Alberts} et~al.,}{{Alberts}
  et~al.}{2013}]{2013MNRAS.431..194A}
{Alberts} S.,  et~al., 2013, \mn@doi [\mnras] {10.1093/mnras/stt155}, \href
  {http://adsabs.harvard.edu/abs/2013MNRAS.431..194A} {431, 194}

\bibitem[\protect\citeauthoryear{{{\'A}lvarez-M{\'a}rquez}
  et~al.,}{{{\'A}lvarez-M{\'a}rquez} et~al.}{2016}]{2016A&A...587A.122A}
{{\'A}lvarez-M{\'a}rquez} J.,  et~al., 2016, \mn@doi [\aap]
  {10.1051/0004-6361/201527190}, \href
  {http://adsabs.harvard.edu/abs/2016A%26A...587A.122A} {587, A122}

\bibitem[\protect\citeauthoryear{{Aretxaga}, {Hughes}, {Chapin},
  {Gazta{\~n}aga}, {Dunlop}  \& {Ivison}}{{Aretxaga}
  et~al.}{2003}]{2003MNRAS.342..759A}
{Aretxaga} I.,  {Hughes} D.~H.,  {Chapin} E.~L.,  {Gazta{\~n}aga} E.,  {Dunlop}
  J.~S.,   {Ivison} R.~J.,  2003, \mn@doi [\mnras]
  {10.1046/j.1365-8711.2003.06560.x}, \href
  {http://adsabs.harvard.edu/abs/2003MNRAS.342..759A} {342, 759}

\bibitem[\protect\citeauthoryear{{Aretxaga} et~al.,}{{Aretxaga}
  et~al.}{2007}]{2007MNRAS.379.1571A}
{Aretxaga} I.,  et~al., 2007, \mn@doi [\mnras]
  {10.1111/j.1365-2966.2007.12036.x}, \href
  {http://adsabs.harvard.edu/abs/2007MNRAS.379.1571A} {379, 1571}

\bibitem[\protect\citeauthoryear{{Ashby} et~al.,}{{Ashby}
  et~al.}{2006}]{2006ApJ...644..778A}
{Ashby} M.~L.~N.,  et~al., 2006, \mn@doi [\apj] {10.1086/503861}, \href
  {http://adsabs.harvard.edu/abs/2006ApJ...644..778A} {644, 778}

\bibitem[\protect\citeauthoryear{{Barger}, {Cowie}, {Sanders}, {Fulton},
  {Taniguchi}, {Sato}, {Kawara}  \& {Okuda}}{{Barger}
  et~al.}{1998}]{1998Natur.394..248B}
{Barger} A.~J.,  {Cowie} L.~L.,  {Sanders} D.~B.,  {Fulton} E.,  {Taniguchi}
  Y.,  {Sato} Y.,  {Kawara} K.,   {Okuda} H.,  1998, \mn@doi [\nat]
  {10.1038/28338}, \href {http://adsabs.harvard.edu/abs/1998Natur.394..248B}
  {394, 248}

\bibitem[\protect\citeauthoryear{{Barger}, {Wang}, {Cowie}, {Owen}, {Chen}  \&
  {Williams}}{{Barger} et~al.}{2012}]{2012ApJ...761...89B}
{Barger} A.~J.,  {Wang} W.-H.,  {Cowie} L.~L.,  {Owen} F.~N.,  {Chen} C.-C.,
  {Williams} J.~P.,  2012, \mn@doi [\apj] {10.1088/0004-637X/761/2/89}, \href
  {http://adsabs.harvard.edu/abs/2012ApJ...761...89B} {761, 89}

\bibitem[\protect\citeauthoryear{{Barger} et~al.,}{{Barger}
  et~al.}{2014}]{2014ApJ...784....9B}
{Barger} A.~J.,  et~al., 2014, \mn@doi [\apj] {10.1088/0004-637X/784/1/9},
  \href {http://adsabs.harvard.edu/abs/2014ApJ...784....9B} {784, 9}

\bibitem[\protect\citeauthoryear{{Barro} et~al.,}{{Barro}
  et~al.}{2011a}]{2011ApJS..193...13B}
{Barro} G.,  et~al., 2011a, \mn@doi [\apjs] {10.1088/0067-0049/193/1/13}, \href
  {http://adsabs.harvard.edu/abs/2011ApJS..193...13B} {193, 13}

\bibitem[\protect\citeauthoryear{{Barro} et~al.,}{{Barro}
  et~al.}{2011b}]{2011ApJS..193...30B}
{Barro} G.,  et~al., 2011b, \mn@doi [\apjs] {10.1088/0067-0049/193/2/30}, \href
  {http://adsabs.harvard.edu/abs/2011ApJS..193...30B} {193, 30}

\bibitem[\protect\citeauthoryear{{Bendo} et~al.,}{{Bendo}
  et~al.}{2013}]{2013MNRAS.433.3062B}
{Bendo} G.~J.,  et~al., 2013, \mn@doi [\mnras] {10.1093/mnras/stt948}, \href
  {http://adsabs.harvard.edu/abs/2013MNRAS.433.3062B} {433, 3062}

\bibitem[\protect\citeauthoryear{{B{\'e}thermin} et~al.,}{{B{\'e}thermin}
  et~al.}{2015a}]{2015A&A...573A.113B}
{B{\'e}thermin} M.,  et~al., 2015a, \mn@doi [\aap]
  {10.1051/0004-6361/201425031}, \href
  {http://adsabs.harvard.edu/abs/2015A%26A...573A.113B} {573, A113}

\bibitem[\protect\citeauthoryear{{B{\'e}thermin}, {De Breuck}, {Sargent}  \&
  {Daddi}}{{B{\'e}thermin} et~al.}{2015b}]{2015A&A...576L...9B}
{B{\'e}thermin} M.,  {De Breuck} C.,  {Sargent} M.,   {Daddi} E.,  2015b,
  \mn@doi [\aap] {10.1051/0004-6361/201525718}, \href
  {http://adsabs.harvard.edu/abs/2015A%26A...576L...9B} {576, L9}

\bibitem[\protect\citeauthoryear{{Biggs} et~al.,}{{Biggs}
  et~al.}{2011}]{2011MNRAS.413.2314B}
{Biggs} A.~D.,  et~al., 2011, \mn@doi [\mnras]
  {10.1111/j.1365-2966.2010.18132.x}, \href
  {http://adsabs.harvard.edu/abs/2011MNRAS.413.2314B} {413, 2314}

\bibitem[\protect\citeauthoryear{{Blain}, {Smail}, {Ivison}, {Kneib}  \&
  {Frayer}}{{Blain} et~al.}{2002}]{2002PhR...369..111B}
{Blain} A.~W.,  {Smail} I.,  {Ivison} R.~J.,  {Kneib} J.-P.,   {Frayer} D.~T.,
  2002, \mn@doi [\physrep] {10.1016/S0370-1573(02)00134-5}, \href
  {http://adsabs.harvard.edu/abs/2002PhR...369..111B} {369, 111}

\bibitem[\protect\citeauthoryear{{Bourne} et~al.,}{{Bourne}
  et~al.}{2014}]{2014MNRAS.444.1884B}
{Bourne} N.,  et~al., 2014, \mn@doi [\mnras] {10.1093/mnras/stu1582}, \href
  {http://adsabs.harvard.edu/abs/2014MNRAS.444.1884B} {444, 1884}

\bibitem[\protect\citeauthoryear{{Bourne} et~al.,}{{Bourne}
  et~al.}{2017}]{2017MNRAS.467.1360B}
{Bourne} N.,  et~al., 2017, \mn@doi [\mnras] {10.1093/mnras/stx031}, \href
  {http://adsabs.harvard.edu/abs/2017MNRAS.467.1360B} {467, 1360}

\bibitem[\protect\citeauthoryear{{Bouwens} et~al.,}{{Bouwens}
  et~al.}{2016}]{2016ApJ...833...72B}
{Bouwens} R.~J.,  et~al., 2016, \mn@doi [\apj] {10.3847/1538-4357/833/1/72},
  \href {http://adsabs.harvard.edu/abs/2016ApJ...833...72B} {833, 72}

\bibitem[\protect\citeauthoryear{{Brammer}, {van Dokkum}  \& {Coppi}}{{Brammer}
  et~al.}{2008}]{2008ApJ...686.1503B}
{Brammer} G.~B.,  {van Dokkum} P.~G.,   {Coppi} P.,  2008, \mn@doi [\apj]
  {10.1086/591786}, \href {http://adsabs.harvard.edu/abs/2008ApJ...686.1503B}
  {686, 1503}

\bibitem[\protect\citeauthoryear{{Brammer} et~al.,}{{Brammer}
  et~al.}{2012}]{2012ApJS..200...13B}
{Brammer} G.~B.,  et~al., 2012, \mn@doi [\apjs] {10.1088/0067-0049/200/2/13},
  \href {http://adsabs.harvard.edu/abs/2012ApJS..200...13B} {200, 13}

\bibitem[\protect\citeauthoryear{{Brisbin} et~al.,}{{Brisbin}
  et~al.}{2017}]{2017A&A...608A..15B}
{Brisbin} D.,  et~al., 2017, \mn@doi [\aap] {10.1051/0004-6361/201730558},
  \href {http://adsabs.harvard.edu/abs/2017A%26A...608A..15B} {608, A15}

\bibitem[\protect\citeauthoryear{{Bruce} et~al.,}{{Bruce}
  et~al.}{2014}]{2014MNRAS.444.1001B}
{Bruce} V.~A.,  et~al., 2014, \mn@doi [\mnras] {10.1093/mnras/stu1478}, \href
  {http://adsabs.harvard.edu/abs/2014MNRAS.444.1001B} {444, 1001}

\bibitem[\protect\citeauthoryear{{Bruzual} \& {Charlot}}{{Bruzual} \&
  {Charlot}}{2003}]{2003MNRAS.344.1000B}
{Bruzual} G.,  {Charlot} S.,  2003, \mn@doi [\mnras]
  {10.1046/j.1365-8711.2003.06897.x}, \href
  {http://adsabs.harvard.edu/abs/2003MNRAS.344.1000B} {344, 1000}

\bibitem[\protect\citeauthoryear{{Calzetti}, {Armus}, {Bohlin}, {Kinney},
  {Koornneef}  \& {Storchi-Bergmann}}{{Calzetti}
  et~al.}{2000}]{2000ApJ...533..682C}
{Calzetti} D.,  {Armus} L.,  {Bohlin} R.~C.,  {Kinney} A.~L.,  {Koornneef} J.,
   {Storchi-Bergmann} T.,  2000, \mn@doi [\apj] {10.1086/308692}, \href
  {http://adsabs.harvard.edu/abs/2000ApJ...533..682C} {533, 682}

\bibitem[\protect\citeauthoryear{{Capak} et~al.,}{{Capak}
  et~al.}{2015}]{2015Natur.522..455C}
{Capak} P.~L.,  et~al., 2015, \mn@doi [\nat] {10.1038/nature14500}, \href
  {http://adsabs.harvard.edu/abs/2015Natur.522..455C} {522, 455}

\bibitem[\protect\citeauthoryear{{Casey} et~al.,}{{Casey}
  et~al.}{2012}]{2012ApJ...761..139C}
{Casey} C.~M.,  et~al., 2012, \mn@doi [\apj] {10.1088/0004-637X/761/2/139},
  \href {http://adsabs.harvard.edu/abs/2012ApJ...761..139C} {761, 139}

\bibitem[\protect\citeauthoryear{{Casey} et~al.,}{{Casey}
  et~al.}{2013}]{2013MNRAS.436.1919C}
{Casey} C.~M.,  et~al., 2013, \mn@doi [\mnras] {10.1093/mnras/stt1673}, \href
  {http://adsabs.harvard.edu/abs/2013MNRAS.436.1919C} {436, 1919}

\bibitem[\protect\citeauthoryear{{Casey}, {Narayanan}  \& {Cooray}}{{Casey}
  et~al.}{2014a}]{2014PhR...541...45C}
{Casey} C.~M.,  {Narayanan} D.,   {Cooray} A.,  2014a, \mn@doi [\physrep]
  {10.1016/j.physrep.2014.02.009}, \href
  {http://adsabs.harvard.edu/abs/2014PhR...541...45C} {541, 45}

\bibitem[\protect\citeauthoryear{{Casey} et~al.,}{{Casey}
  et~al.}{2014b}]{2014ApJ...796...95C}
{Casey} C.~M.,  et~al., 2014b, \mn@doi [\apj] {10.1088/0004-637X/796/2/95},
  \href {http://adsabs.harvard.edu/abs/2014ApJ...796...95C} {796, 95}

\bibitem[\protect\citeauthoryear{{Chabrier}}{{Chabrier}}{2003}]{2003PASP..115..763C}
{Chabrier} G.,  2003, \mn@doi [\pasp] {10.1086/376392}, \href
  {http://adsabs.harvard.edu/abs/2003PASP..115..763C} {115, 763}

\bibitem[\protect\citeauthoryear{{Chapin}, {Hughes}  \& {Aretxaga}}{{Chapin}
  et~al.}{2009a}]{2009MNRAS.393..653C}
{Chapin} E.~L.,  {Hughes} D.~H.,   {Aretxaga} I.,  2009a, \mn@doi [\mnras]
  {10.1111/j.1365-2966.2008.14242.x}, \href
  {http://adsabs.harvard.edu/abs/2009MNRAS.393..653C} {393, 653}

\bibitem[\protect\citeauthoryear{{Chapin} et~al.,}{{Chapin}
  et~al.}{2009b}]{2009MNRAS.398.1793C}
{Chapin} E.~L.,  et~al., 2009b, \mn@doi [\mnras]
  {10.1111/j.1365-2966.2009.15267.x}, \href
  {http://adsabs.harvard.edu/abs/2009MNRAS.398.1793C} {398, 1793}

\bibitem[\protect\citeauthoryear{{Chapin} et~al.,}{{Chapin}
  et~al.}{2011}]{2011MNRAS.411..505C}
{Chapin} E.~L.,  et~al., 2011, \mn@doi [\mnras]
  {10.1111/j.1365-2966.2010.17697.x}, \href
  {http://adsabs.harvard.edu/abs/2011MNRAS.411..505C} {411, 505}

\bibitem[\protect\citeauthoryear{{Chapman}, {Blain}, {Smail}  \&
  {Ivison}}{{Chapman} et~al.}{2005}]{2005ApJ...622..772C}
{Chapman} S.~C.,  {Blain} A.~W.,  {Smail} I.,   {Ivison} R.~J.,  2005, \mn@doi
  [\apj] {10.1086/428082}, \href
  {http://adsabs.harvard.edu/abs/2005ApJ...622..772C} {622, 772}

\bibitem[\protect\citeauthoryear{{Chen}, {Cowie}, {Barger}, {Casey}, {Lee},
  {Sanders}, {Wang}  \& {Williams}}{{Chen} et~al.}{2013}]{2013ApJ...776..131C}
{Chen} C.-C.,  {Cowie} L.~L.,  {Barger} A.~J.,  {Casey} C.~M.,  {Lee} N.,
  {Sanders} D.~B.,  {Wang} W.-H.,   {Williams} J.~P.,  2013, \mn@doi [\apj]
  {10.1088/0004-637X/776/2/131}, \href
  {http://adsabs.harvard.edu/abs/2013ApJ...776..131C} {776, 131}

\bibitem[\protect\citeauthoryear{{Chen} et~al.,}{{Chen}
  et~al.}{2015}]{2015ApJ...799..194C}
{Chen} C.-C.,  et~al., 2015, \mn@doi [\apj] {10.1088/0004-637X/799/2/194},
  \href {http://adsabs.harvard.edu/abs/2015ApJ...799..194C} {799, 194}

\bibitem[\protect\citeauthoryear{{Chen} et~al.,}{{Chen}
  et~al.}{2016}]{2016ApJ...820...82C}
{Chen} C.-C.,  et~al., 2016, \mn@doi [\apj] {10.3847/0004-637X/820/2/82}, \href
  {http://adsabs.harvard.edu/abs/2016ApJ...820...82C} {820, 82}

\bibitem[\protect\citeauthoryear{{Cowie}, {Barger}, {Hsu}, {Chen}, {Owen}  \&
  {Wang}}{{Cowie} et~al.}{2017}]{2017ApJ...837..139C}
{Cowie} L.~L.,  {Barger} A.~J.,  {Hsu} L.-Y.,  {Chen} C.-C.,  {Owen} F.~N.,
  {Wang} W.-H.,  2017, \mn@doi [\apj] {10.3847/1538-4357/aa60bb}, \href
  {http://adsabs.harvard.edu/abs/2017ApJ...837..139C} {837, 139}

\bibitem[\protect\citeauthoryear{{Cowley}, {B{\'e}thermin}, {del P.~Lagos},
  {Lacey}, {Baugh}  \& {Cole}}{{Cowley} et~al.}{2017}]{2017MNRAS.tmp..172C}
{Cowley} W.~I.,  {B{\'e}thermin} M.,  {del P.~Lagos} C.,  {Lacey} C.~G.,
  {Baugh} C.~M.,   {Cole} S.,  2017, \mn@doi [\mnras] {10.1093/mnras/stx165},
  \href {http://adsabs.harvard.edu/abs/2017MNRAS.tmp..172C} {}

\bibitem[\protect\citeauthoryear{{Dav{\'e}}, {Finlator}, {Oppenheimer},
  {Fardal}, {Katz}, {Kere{\v s}}  \& {Weinberg}}{{Dav{\'e}}
  et~al.}{2010}]{2010MNRAS.404.1355D}
{Dav{\'e}} R.,  {Finlator} K.,  {Oppenheimer} B.~D.,  {Fardal} M.,  {Katz} N.,
  {Kere{\v s}} D.,   {Weinberg} D.~H.,  2010, \mn@doi [\mnras]
  {10.1111/j.1365-2966.2010.16395.x}, \href
  {http://adsabs.harvard.edu/abs/2010MNRAS.404.1355D} {404, 1355}

\bibitem[\protect\citeauthoryear{{Dempsey} et~al.,}{{Dempsey}
  et~al.}{2013}]{2013MNRAS.430.2534D}
{Dempsey} J.~T.,  et~al., 2013, \mn@doi [\mnras] {10.1093/mnras/stt090}, \href
  {http://adsabs.harvard.edu/abs/2013MNRAS.430.2534D} {430, 2534}

\bibitem[\protect\citeauthoryear{{Downes}, {Peacock}, {Savage}  \&
  {Carrie}}{{Downes} et~al.}{1986}]{1986MNRAS.218...31D}
{Downes} A.~J.~B.,  {Peacock} J.~A.,  {Savage} A.,   {Carrie} D.~R.,  1986,
  \mn@doi [\mnras] {10.1093/mnras/218.1.31}, \href
  {http://adsabs.harvard.edu/abs/1986MNRAS.218...31D} {218, 31}

\bibitem[\protect\citeauthoryear{{Dunlop}, {Peacock}, {Savage}, {Lilly},
  {Heasley}  \& {Simon}}{{Dunlop} et~al.}{1989}]{1989MNRAS.238.1171D}
{Dunlop} J.~S.,  {Peacock} J.~A.,  {Savage} A.,  {Lilly} S.~J.,  {Heasley}
  J.~N.,   {Simon} A.~J.~B.,  1989, \mn@doi [\mnras]
  {10.1093/mnras/238.4.1171}, \href
  {http://adsabs.harvard.edu/abs/1989MNRAS.238.1171D} {238, 1171}

\bibitem[\protect\citeauthoryear{{Dunlop} et~al.,}{{Dunlop}
  et~al.}{2017}]{2017MNRAS.466..861D}
{Dunlop} J.~S.,  et~al., 2017, \mn@doi [\mnras] {10.1093/mnras/stw3088}, \href
  {http://adsabs.harvard.edu/abs/2017MNRAS.466..861D} {466, 861}

\bibitem[\protect\citeauthoryear{{Dunne} \& {Eales}}{{Dunne} \&
  {Eales}}{2001}]{2001MNRAS.327..697D}
{Dunne} L.,  {Eales} S.~A.,  2001, \mn@doi [\mnras]
  {10.1046/j.1365-8711.2001.04789.x}, \href
  {http://adsabs.harvard.edu/abs/2001MNRAS.327..697D} {327, 697}

\bibitem[\protect\citeauthoryear{{Farrah}, {Afonso}, {Efstathiou},
  {Rowan-Robinson}, {Fox}  \& {Clements}}{{Farrah}
  et~al.}{2003}]{2003MNRAS.343..585F}
{Farrah} D.,  {Afonso} J.,  {Efstathiou} A.,  {Rowan-Robinson} M.,  {Fox} M.,
  {Clements} D.,  2003, \mn@doi [\mnras] {10.1046/j.1365-8711.2003.06696.x},
  \href {http://adsabs.harvard.edu/abs/2003MNRAS.343..585F} {343, 585}

\bibitem[\protect\citeauthoryear{{Fixsen}, {Dwek}, {Mather}, {Bennett}  \&
  {Shafer}}{{Fixsen} et~al.}{1998}]{1998ApJ...508..123F}
{Fixsen} D.~J.,  {Dwek} E.,  {Mather} J.~C.,  {Bennett} C.~L.,   {Shafer}
  R.~A.,  1998, \mn@doi [\apj] {10.1086/306383}, \href
  {http://adsabs.harvard.edu/abs/1998ApJ...508..123F} {508, 123}

\bibitem[\protect\citeauthoryear{{Geach} et~al.,}{{Geach}
  et~al.}{2013}]{2013MNRAS.432...53G}
{Geach} J.~E.,  et~al., 2013, \mn@doi [\mnras] {10.1093/mnras/stt352}, \href
  {http://adsabs.harvard.edu/abs/2013MNRAS.432...53G} {432, 53}

\bibitem[\protect\citeauthoryear{{Geach} et~al.,}{{Geach}
  et~al.}{2015}]{2015MNRAS.452..502G}
{Geach} J.~E.,  et~al., 2015, \mn@doi [\mnras] {10.1093/mnras/stv1243}, \href
  {http://adsabs.harvard.edu/abs/2015MNRAS.452..502G} {452, 502}

\bibitem[\protect\citeauthoryear{{Geach} et~al.,}{{Geach}
  et~al.}{2017}]{2017MNRAS.465.1789G}
{Geach} J.~E.,  et~al., 2017, \mn@doi [\mnras] {10.1093/mnras/stw2721}, \href
  {http://adsabs.harvard.edu/abs/2017MNRAS.465.1789G} {465, 1789}

\bibitem[\protect\citeauthoryear{{Genzel} et~al.,}{{Genzel}
  et~al.}{2015}]{2015ApJ...800...20G}
{Genzel} R.,  et~al., 2015, \mn@doi [\apj] {10.1088/0004-637X/800/1/20}, \href
  {http://adsabs.harvard.edu/abs/2015ApJ...800...20G} {800, 20}

\bibitem[\protect\citeauthoryear{{Greve} et~al.,}{{Greve}
  et~al.}{2005}]{2005MNRAS.359.1165G}
{Greve} T.~R.,  et~al., 2005, \mn@doi [\mnras]
  {10.1111/j.1365-2966.2005.08979.x}, \href
  {http://adsabs.harvard.edu/abs/2005MNRAS.359.1165G} {359, 1165}

\bibitem[\protect\citeauthoryear{{Griffin} et~al.,}{{Griffin}
  et~al.}{2010}]{2010A&A...518L...3G}
{Griffin} M.~J.,  et~al., 2010, \mn@doi [\aap] {10.1051/0004-6361/201014519},
  \href {http://adsabs.harvard.edu/abs/2010A%26A...518L...3G} {518, L3}

\bibitem[\protect\citeauthoryear{{Grogin} et~al.,}{{Grogin}
  et~al.}{2011}]{2011ApJS..197...35G}
{Grogin} N.~A.,  et~al., 2011, \mn@doi [\apjs] {10.1088/0067-0049/197/2/35},
  \href {http://adsabs.harvard.edu/abs/2011ApJS..197...35G} {197, 35}

\bibitem[\protect\citeauthoryear{{Gruppioni} et~al.,}{{Gruppioni}
  et~al.}{2013}]{2013MNRAS.432...23G}
{Gruppioni} C.,  et~al., 2013, \mn@doi [\mnras] {10.1093/mnras/stt308}, \href
  {http://adsabs.harvard.edu/abs/2013MNRAS.432...23G} {432, 23}

\bibitem[\protect\citeauthoryear{{Hatsukade} et~al.,}{{Hatsukade}
  et~al.}{2016}]{2016PASJ...68...36H}
{Hatsukade} B.,  et~al., 2016, \mn@doi [\pasj] {10.1093/pasj/psw026}, \href
  {http://adsabs.harvard.edu/abs/2016PASJ...68...36H} {68, 36}

\bibitem[\protect\citeauthoryear{{Hickox} et~al.,}{{Hickox}
  et~al.}{2012}]{2012MNRAS.421..284H}
{Hickox} R.~C.,  et~al., 2012, \mn@doi [\mnras]
  {10.1111/j.1365-2966.2011.20303.x}, \href
  {http://adsabs.harvard.edu/abs/2012MNRAS.421..284H} {421, 284}

\bibitem[\protect\citeauthoryear{{Hodge} et~al.,}{{Hodge}
  et~al.}{2013}]{2013ApJ...768...91H}
{Hodge} J.~A.,  et~al., 2013, \mn@doi [\apj] {10.1088/0004-637X/768/1/91},
  \href {http://adsabs.harvard.edu/abs/2013ApJ...768...91H} {768, 91}

\bibitem[\protect\citeauthoryear{{Hodge} et~al.,}{{Hodge}
  et~al.}{2016}]{2016ApJ...833..103H}
{Hodge} J.~A.,  et~al., 2016, \mn@doi [\apj] {10.3847/1538-4357/833/1/103},
  \href {http://adsabs.harvard.edu/abs/2016ApJ...833..103H} {833, 103}

\bibitem[\protect\citeauthoryear{{Holland} et~al.,}{{Holland}
  et~al.}{2013}]{2013MNRAS.430.2513H}
{Holland} W.~S.,  et~al., 2013, \mn@doi [\mnras] {10.1093/mnras/sts612}, \href
  {http://adsabs.harvard.edu/abs/2013MNRAS.430.2513H} {430, 2513}

\bibitem[\protect\citeauthoryear{{Hsu}, {Cowie}, {Chen}, {Barger}  \&
  {Wang}}{{Hsu} et~al.}{2016}]{2016arXiv160500046H}
{Hsu} L.-Y.,  {Cowie} L.,  {Chen} C.-C.,  {Barger} A.,   {Wang} W.-H.,  2016,
  preprint, \href {http://adsabs.harvard.edu/abs/2016arXiv160500046H} {}
  (\mn@eprint {arXiv} {1605.00046})

\bibitem[\protect\citeauthoryear{{Huertas-Company} et~al.,}{{Huertas-Company}
  et~al.}{2015a}]{2015ApJS..221....8H}
{Huertas-Company} M.,  et~al., 2015a, \mn@doi [\apjs]
  {10.1088/0067-0049/221/1/8}, \href
  {http://adsabs.harvard.edu/abs/2015ApJS..221....8H} {221, 8}

\bibitem[\protect\citeauthoryear{{Huertas-Company} et~al.,}{{Huertas-Company}
  et~al.}{2015b}]{2015ApJ...809...95H}
{Huertas-Company} M.,  et~al., 2015b, \mn@doi [\apj]
  {10.1088/0004-637X/809/1/95}, \href
  {http://adsabs.harvard.edu/abs/2015ApJ...809...95H} {809, 95}

\bibitem[\protect\citeauthoryear{{Hughes} et~al.,}{{Hughes}
  et~al.}{1998}]{1998Natur.394..241H}
{Hughes} D.~H.,  et~al., 1998, \mn@doi [\nat] {10.1038/28328}, \href
  {http://adsabs.harvard.edu/abs/1998Natur.394..241H} {394, 241}

\bibitem[\protect\citeauthoryear{{Ikarashi} et~al.,}{{Ikarashi}
  et~al.}{2015}]{2015ApJ...810..133I}
{Ikarashi} S.,  et~al., 2015, \mn@doi [\apj] {10.1088/0004-637X/810/2/133},
  \href {http://adsabs.harvard.edu/abs/2015ApJ...810..133I} {810, 133}

\bibitem[\protect\citeauthoryear{{Iono} et~al.,}{{Iono}
  et~al.}{2006}]{2006ApJ...640L...1I}
{Iono} D.,  et~al., 2006, \mn@doi [\apjl] {10.1086/503290}, \href
  {http://adsabs.harvard.edu/abs/2006ApJ...640L...1I} {640, L1}

\bibitem[\protect\citeauthoryear{{Isobe}, {Feigelson}  \& {Nelson}}{{Isobe}
  et~al.}{1986}]{1986ApJ...306..490I}
{Isobe} T.,  {Feigelson} E.~D.,   {Nelson} P.~I.,  1986, \mn@doi [\apj]
  {10.1086/164359}, \href {http://adsabs.harvard.edu/abs/1986ApJ...306..490I}
  {306, 490}

\bibitem[\protect\citeauthoryear{{Ivison} et~al.,}{{Ivison}
  et~al.}{2007a}]{2007MNRAS.380..199I}
{Ivison} R.~J.,  et~al., 2007a, \mn@doi [\mnras]
  {10.1111/j.1365-2966.2007.12044.x}, \href
  {http://adsabs.harvard.edu/abs/2007MNRAS.380..199I} {380, 199}

\bibitem[\protect\citeauthoryear{{Ivison} et~al.,}{{Ivison}
  et~al.}{2007b}]{2007ApJ...660L..77I}
{Ivison} R.~J.,  et~al., 2007b, \mn@doi [\apjl] {10.1086/517917}, \href
  {http://adsabs.harvard.edu/abs/2007ApJ...660L..77I} {660, L77}

\bibitem[\protect\citeauthoryear{{Ivison} et~al.,}{{Ivison}
  et~al.}{2016}]{2016ApJ...832...78I}
{Ivison} R.~J.,  et~al., 2016, \mn@doi [\apj] {10.3847/0004-637X/832/1/78},
  \href {http://adsabs.harvard.edu/abs/2016ApJ...832...78I} {832, 78}

\bibitem[\protect\citeauthoryear{{Karim} et~al.,}{{Karim}
  et~al.}{2013}]{2013MNRAS.432....2K}
{Karim} A.,  et~al., 2013, \mn@doi [\mnras] {10.1093/mnras/stt196}, \href
  {http://adsabs.harvard.edu/abs/2013MNRAS.432....2K} {432, 2}

\bibitem[\protect\citeauthoryear{{Kartaltepe} et~al.,}{{Kartaltepe}
  et~al.}{2015}]{2015ApJS..221...11K}
{Kartaltepe} J.~S.,  et~al., 2015, \mn@doi [\apjs]
  {10.1088/0067-0049/221/1/11}, \href
  {http://adsabs.harvard.edu/abs/2015ApJS..221...11K} {221, 11}

\bibitem[\protect\citeauthoryear{{Kennicutt}}{{Kennicutt}}{1998}]{1998ARA&A..36..189K}
{Kennicutt} Jr. R.~C.,  1998, \mn@doi [\araa] {10.1146/annurev.astro.36.1.189},
  \href {http://adsabs.harvard.edu/abs/1998ARA%26A..36..189K} {36, 189}

\bibitem[\protect\citeauthoryear{{Kirkpatrick}, {Pope}, {Sajina}, {Roebuck},
  {Yan}, {Armus}, {D{\'{\i}}az-Santos}  \& {Stierwalt}}{{Kirkpatrick}
  et~al.}{2015}]{2015ApJ...814....9K}
{Kirkpatrick} A.,  {Pope} A.,  {Sajina} A.,  {Roebuck} E.,  {Yan} L.,  {Armus}
  L.,  {D{\'{\i}}az-Santos} T.,   {Stierwalt} S.,  2015, \mn@doi [\apj]
  {10.1088/0004-637X/814/1/9}, \href
  {http://adsabs.harvard.edu/abs/2015ApJ...814....9K} {814, 9}

\bibitem[\protect\citeauthoryear{{Kong}, {Charlot}, {Brinchmann}  \&
  {Fall}}{{Kong} et~al.}{2004}]{2004MNRAS.349..769K}
{Kong} X.,  {Charlot} S.,  {Brinchmann} J.,   {Fall} S.~M.,  2004, \mn@doi
  [\mnras] {10.1111/j.1365-2966.2004.07556.x}, \href
  {http://adsabs.harvard.edu/abs/2004MNRAS.349..769K} {349, 769}

\bibitem[\protect\citeauthoryear{{Koprowski}, {Dunlop}, {Micha{\l}owski},
  {Cirasuolo}  \& {Bowler}}{{Koprowski} et~al.}{2014}]{2014MNRAS.444..117K}
{Koprowski} M.~P.,  {Dunlop} J.~S.,  {Micha{\l}owski} M.~J.,  {Cirasuolo} M.,
  {Bowler} R.~A.~A.,  2014, \mn@doi [\mnras] {10.1093/mnras/stu1402}, \href
  {http://adsabs.harvard.edu/abs/2014MNRAS.444..117K} {444, 117}

\bibitem[\protect\citeauthoryear{{Koprowski} et~al.,}{{Koprowski}
  et~al.}{2016}]{2016MNRAS.458.4321K}
{Koprowski} M.~P.,  et~al., 2016, \mn@doi [\mnras] {10.1093/mnras/stw564},
  \href {http://adsabs.harvard.edu/abs/2016MNRAS.458.4321K} {458, 4321}

\bibitem[\protect\citeauthoryear{{Kov{\'a}cs}, {Chapman}, {Dowell}, {Blain},
  {Ivison}, {Smail}  \& {Phillips}}{{Kov{\'a}cs}
  et~al.}{2006}]{2006ApJ...650..592K}
{Kov{\'a}cs} A.,  {Chapman} S.~C.,  {Dowell} C.~D.,  {Blain} A.~W.,  {Ivison}
  R.~J.,  {Smail} I.,   {Phillips} T.~G.,  2006, \mn@doi [\apj]
  {10.1086/506341}, \href {http://adsabs.harvard.edu/abs/2006ApJ...650..592K}
  {650, 592}

\bibitem[\protect\citeauthoryear{{Kriek}, {van Dokkum}, {Labb{\'e}}, {Franx},
  {Illingworth}, {Marchesini}  \& {Quadri}}{{Kriek}
  et~al.}{2009}]{2009ApJ...700..221K}
{Kriek} M.,  {van Dokkum} P.~G.,  {Labb{\'e}} I.,  {Franx} M.,  {Illingworth}
  G.~D.,  {Marchesini} D.,   {Quadri} R.~F.,  2009, \mn@doi [\apj]
  {10.1088/0004-637X/700/1/221}, \href
  {http://adsabs.harvard.edu/abs/2009ApJ...700..221K} {700, 221}

\bibitem[\protect\citeauthoryear{{Lacey} et~al.,}{{Lacey}
  et~al.}{2016}]{2016MNRAS.462.3854L}
{Lacey} C.~G.,  et~al., 2016, \mn@doi [\mnras] {10.1093/mnras/stw1888}, \href
  {http://adsabs.harvard.edu/abs/2016MNRAS.462.3854L} {462, 3854}

\bibitem[\protect\citeauthoryear{{Lagos}, {Bayet}, {Baugh}, {Lacey}, {Bell},
  {Fanidakis}  \& {Geach}}{{Lagos} et~al.}{2012}]{2012MNRAS.426.2142L}
{Lagos} C.~d.~P.,  {Bayet} E.,  {Baugh} C.~M.,  {Lacey} C.~G.,  {Bell} T.~A.,
  {Fanidakis} N.,   {Geach} J.~E.,  2012, \mn@doi [\mnras]
  {10.1111/j.1365-2966.2012.21905.x}, \href
  {http://adsabs.harvard.edu/abs/2012MNRAS.426.2142L} {426, 2142}

\bibitem[\protect\citeauthoryear{{Laird} et~al.,}{{Laird}
  et~al.}{2009}]{2009ApJS..180..102L}
{Laird} E.~S.,  et~al., 2009, \mn@doi [\apjs] {10.1088/0067-0049/180/1/102},
  \href {http://adsabs.harvard.edu/abs/2009ApJS..180..102L} {180, 102}

\bibitem[\protect\citeauthoryear{{Lilly}, {Eales}, {Gear}, {Hammer}, {Le
  F{\`e}vre}, {Crampton}, {Bond}  \& {Dunne}}{{Lilly}
  et~al.}{1999}]{1999ApJ...518..641L}
{Lilly} S.~J.,  {Eales} S.~A.,  {Gear} W.~K.~P.,  {Hammer} F.,  {Le F{\`e}vre}
  O.,  {Crampton} D.,  {Bond} J.~R.,   {Dunne} L.,  1999, \mn@doi [\apj]
  {10.1086/307310}, \href {http://adsabs.harvard.edu/abs/1999ApJ...518..641L}
  {518, 641}

\bibitem[\protect\citeauthoryear{{Lutz} et~al.,}{{Lutz}
  et~al.}{2011}]{2011A&A...532A..90L}
{Lutz} D.,  et~al., 2011, \mn@doi [\aap] {10.1051/0004-6361/201117107}, \href
  {http://adsabs.harvard.edu/abs/2011A%26A...532A..90L} {532, A90}

\bibitem[\protect\citeauthoryear{{Magdis} et~al.,}{{Magdis}
  et~al.}{2012}]{2012ApJ...760....6M}
{Magdis} G.~E.,  et~al., 2012, \mn@doi [\apj] {10.1088/0004-637X/760/1/6},
  \href {http://adsabs.harvard.edu/abs/2012ApJ...760....6M} {760, 6}

\bibitem[\protect\citeauthoryear{{Magnelli} et~al.,}{{Magnelli}
  et~al.}{2012}]{2012A&A...539A.155M}
{Magnelli} B.,  et~al., 2012, \mn@doi [\aap] {10.1051/0004-6361/201118312},
  \href {http://adsabs.harvard.edu/abs/2012A%26A...539A.155M} {539, A155}

\bibitem[\protect\citeauthoryear{{Magnelli} et~al.,}{{Magnelli}
  et~al.}{2014}]{2014A&A...561A..86M}
{Magnelli} B.,  et~al., 2014, \mn@doi [\aap] {10.1051/0004-6361/201322217},
  \href {http://adsabs.harvard.edu/abs/2014A%26A...561A..86M} {561, A86}

\bibitem[\protect\citeauthoryear{{Mancini}, {Schneider}, {Graziani},
  {Valiante}, {Dayal}, {Maio}  \& {Ciardi}}{{Mancini}
  et~al.}{2016}]{2016MNRAS.462.3130M}
{Mancini} M.,  {Schneider} R.,  {Graziani} L.,  {Valiante} R.,  {Dayal} P.,
  {Maio} U.,   {Ciardi} B.,  2016, \mn@doi [\mnras] {10.1093/mnras/stw1783},
  \href {http://adsabs.harvard.edu/abs/2016MNRAS.462.3130M} {462, 3130}

\bibitem[\protect\citeauthoryear{{Markwardt}}{{Markwardt}}{2009}]{mpfit}
{Markwardt} C.~B.,  2009, ASPC, \href
  {http://adsabs.harvard.edu/abs/2009ASPC..411..251M} {411, 251}

\bibitem[\protect\citeauthoryear{{Marsden} et~al.,}{{Marsden}
  et~al.}{2011}]{2011MNRAS.417.1192M}
{Marsden} G.,  et~al., 2011, \mn@doi [\mnras]
  {10.1111/j.1365-2966.2011.19336.x}, \href
  {http://adsabs.harvard.edu/abs/2011MNRAS.417.1192M} {417, 1192}

\bibitem[\protect\citeauthoryear{{Meurer}, {Heckman}  \& {Calzetti}}{{Meurer}
  et~al.}{1999}]{1999ApJ...521...64M}
{Meurer} G.~R.,  {Heckman} T.~M.,   {Calzetti} D.,  1999, \mn@doi [\apj]
  {10.1086/307523}, \href {http://adsabs.harvard.edu/abs/1999ApJ...521...64M}
  {521, 64}

\bibitem[\protect\citeauthoryear{{Micha{\l}owski}, {Hjorth}  \&
  {Watson}}{{Micha{\l}owski} et~al.}{2010}]{2010A&A...514A..67M}
{Micha{\l}owski} M.,  {Hjorth} J.,   {Watson} D.,  2010, \mn@doi [\aap]
  {10.1051/0004-6361/200913634}, \href
  {http://adsabs.harvard.edu/abs/2010A%26A...514A..67M} {514, A67}

\bibitem[\protect\citeauthoryear{{Micha{\l}owski} et~al.,}{{Micha{\l}owski}
  et~al.}{2012a}]{2012MNRAS.426.1845M}
{Micha{\l}owski} M.~J.,  et~al., 2012a, \mn@doi [\mnras]
  {10.1111/j.1365-2966.2012.21828.x}, \href
  {http://adsabs.harvard.edu/abs/2012MNRAS.426.1845M} {426, 1845}

\bibitem[\protect\citeauthoryear{{Micha{\l}owski}, {Dunlop}, {Cirasuolo},
  {Hjorth}, {Hayward}  \& {Watson}}{{Micha{\l}owski}
  et~al.}{2012b}]{2012A&A...541A..85M}
{Micha{\l}owski} M.~J.,  {Dunlop} J.~S.,  {Cirasuolo} M.,  {Hjorth} J.,
  {Hayward} C.~C.,   {Watson} D.,  2012b, \mn@doi [\aap]
  {10.1051/0004-6361/201016308}, \href
  {http://adsabs.harvard.edu/abs/2012A%26A...541A..85M} {541, A85}

\bibitem[\protect\citeauthoryear{{Micha{\l}owski} et~al.,}{{Micha{\l}owski}
  et~al.}{2017}]{2017MNRAS.469..492M}
{Micha{\l}owski} M.~J.,  et~al., 2017, \mn@doi [\mnras] {10.1093/mnras/stx861},
  \href {http://adsabs.harvard.edu/abs/2017MNRAS.469..492M} {469, 492}

\bibitem[\protect\citeauthoryear{{Miettinen} et~al.,}{{Miettinen}
  et~al.}{2015}]{2015A&A...577A..29M}
{Miettinen} O.,  et~al., 2015, \mn@doi [\aap] {10.1051/0004-6361/201425032},
  \href {http://adsabs.harvard.edu/abs/2015A%26A...577A..29M} {577, A29}

\bibitem[\protect\citeauthoryear{{Momcheva} et~al.,}{{Momcheva}
  et~al.}{2016}]{2016ApJS..225...27M}
{Momcheva} I.~G.,  et~al., 2016, \mn@doi [\apjs] {10.3847/0067-0049/225/2/27},
  \href {http://adsabs.harvard.edu/abs/2016ApJS..225...27M} {225, 27}

\bibitem[\protect\citeauthoryear{{Mortlock} et~al.,}{{Mortlock}
  et~al.}{2013}]{2013MNRAS.433.1185M}
{Mortlock} A.,  et~al., 2013, \mn@doi [\mnras] {10.1093/mnras/stt793}, \href
  {http://adsabs.harvard.edu/abs/2013MNRAS.433.1185M} {433, 1185}

\bibitem[\protect\citeauthoryear{{Nandra} et~al.,}{{Nandra}
  et~al.}{2015}]{2015ApJS..220...10N}
{Nandra} K.,  et~al., 2015, \mn@doi [\apjs] {10.1088/0067-0049/220/1/10}, \href
  {http://adsabs.harvard.edu/abs/2015ApJS..220...10N} {220, 10}

\bibitem[\protect\citeauthoryear{{Narayanan} et~al.,}{{Narayanan}
  et~al.}{2015}]{2015Natur.525..496N}
{Narayanan} D.,  et~al., 2015, \mn@doi [\nat] {10.1038/nature15383}, \href
  {http://adsabs.harvard.edu/abs/2015Natur.525..496N} {525, 496}

\bibitem[\protect\citeauthoryear{{Nguyen} et~al.,}{{Nguyen}
  et~al.}{2010}]{nguyen10}
{Nguyen} H.~T.,  et~al., 2010, \mn@doi [\aap] {10.1051/0004-6361/201014680},
  \href {http://adsabs.harvard.edu/abs/2010A%26A...518L...5N} {518, L5}

\bibitem[\protect\citeauthoryear{{Oliver} et~al.,}{{Oliver}
  et~al.}{2012}]{2012MNRAS.424.1614O}
{Oliver} S.~J.,  et~al., 2012, \mn@doi [\mnras]
  {10.1111/j.1365-2966.2012.20912.x}, \href
  {http://adsabs.harvard.edu/abs/2012MNRAS.424.1614O} {424, 1614}

\bibitem[\protect\citeauthoryear{{Oteo} et~al.,}{{Oteo}
  et~al.}{2013}]{2013A&A...554L...3O}
{Oteo} I.,  et~al., 2013, \mn@doi [\aap] {10.1051/0004-6361/201321478}, \href
  {http://adsabs.harvard.edu/abs/2013A%26A...554L...3O} {554, L3}

\bibitem[\protect\citeauthoryear{{Penner} et~al.,}{{Penner}
  et~al.}{2012}]{2012ApJ...759...28P}
{Penner} K.,  et~al., 2012, \mn@doi [\apj] {10.1088/0004-637X/759/1/28}, \href
  {http://adsabs.harvard.edu/abs/2012ApJ...759...28P} {759, 28}

\bibitem[\protect\citeauthoryear{{Pettini}, {Kellogg}, {Steidel}, {Dickinson},
  {Adelberger}  \& {Giavalisco}}{{Pettini} et~al.}{1998}]{1998ApJ...508..539P}
{Pettini} M.,  {Kellogg} M.,  {Steidel} C.~C.,  {Dickinson} M.,  {Adelberger}
  K.~L.,   {Giavalisco} M.,  1998, \mn@doi [\apj] {10.1086/306431}, \href
  {http://adsabs.harvard.edu/abs/1998ApJ...508..539P} {508, 539}

\bibitem[\protect\citeauthoryear{{Planck Collaboration} et~al.,}{{Planck
  Collaboration} et~al.}{2014}]{2014A&A...571A..16P}
{Planck Collaboration} et~al., 2014, \mn@doi [\aap]
  {10.1051/0004-6361/201321591}, \href
  {http://adsabs.harvard.edu/abs/2014A%26A...571A..16P} {571, A16}

\bibitem[\protect\citeauthoryear{{Poglitsch} et~al.,}{{Poglitsch}
  et~al.}{2010}]{2010A&A...518L...2P}
{Poglitsch} A.,  et~al., 2010, \mn@doi [\aap] {10.1051/0004-6361/201014535},
  \href {http://adsabs.harvard.edu/abs/2010A%26A...518L...2P} {518, L2}

\bibitem[\protect\citeauthoryear{{Pope}, {Borys}, {Scott}, {Conselice},
  {Dickinson}  \& {Mobasher}}{{Pope} et~al.}{2005}]{2005MNRAS.358..149P}
{Pope} A.,  {Borys} C.,  {Scott} D.,  {Conselice} C.,  {Dickinson} M.,
  {Mobasher} B.,  2005, \mn@doi [\mnras] {10.1111/j.1365-2966.2005.08759.x},
  \href {http://adsabs.harvard.edu/abs/2005MNRAS.358..149P} {358, 149}

\bibitem[\protect\citeauthoryear{{Pope} et~al.,}{{Pope}
  et~al.}{2006}]{2006MNRAS.370.1185P}
{Pope} A.,  et~al., 2006, \mn@doi [\mnras] {10.1111/j.1365-2966.2006.10575.x},
  \href {http://adsabs.harvard.edu/abs/2006MNRAS.370.1185P} {370, 1185}

\bibitem[\protect\citeauthoryear{{Pope} et~al.,}{{Pope}
  et~al.}{2017}]{2017arXiv170304535P}
{Pope} A.,  et~al., 2017, preprint, \href
  {http://adsabs.harvard.edu/abs/2017arXiv170304535P} {} (\mn@eprint {arXiv}
  {1703.04535})

\bibitem[\protect\citeauthoryear{{Popping}, {Puglisi}  \& {Norman}}{{Popping}
  et~al.}{2017}]{2017MNRAS.472.2315P}
{Popping} G.,  {Puglisi} A.,   {Norman} C.~A.,  2017, \mn@doi [\mnras]
  {10.1093/mnras/stx2202}, \href
  {http://adsabs.harvard.edu/abs/2017MNRAS.472.2315P} {472, 2315}

\bibitem[\protect\citeauthoryear{{Puget}, {Abergel}, {Bernard}, {Boulanger},
  {Burton}, {Desert}  \& {Hartmann}}{{Puget}
  et~al.}{1996}]{1996A&A...308L...5P}
{Puget} J.-L.,  {Abergel} A.,  {Bernard} J.-P.,  {Boulanger} F.,  {Burton}
  W.~B.,  {Desert} F.-X.,   {Hartmann} D.,  1996, \aap, \href
  {http://adsabs.harvard.edu/abs/1996A%26A...308L...5P} {308, L5}

\bibitem[\protect\citeauthoryear{{Reddy} et~al.,}{{Reddy}
  et~al.}{2012}]{2012ApJ...744..154R}
{Reddy} N.,  et~al., 2012, \mn@doi [\apj] {10.1088/0004-637X/744/2/154}, \href
  {http://adsabs.harvard.edu/abs/2012ApJ...744..154R} {744, 154}

\bibitem[\protect\citeauthoryear{{Riechers} et~al.,}{{Riechers}
  et~al.}{2013}]{2013Natur.496..329R}
{Riechers} D.~A.,  et~al., 2013, \mn@doi [\nat] {10.1038/nature12050}, \href
  {http://adsabs.harvard.edu/abs/2013Natur.496..329R} {496, 329}

\bibitem[\protect\citeauthoryear{{Roseboom} et~al.,}{{Roseboom}
  et~al.}{2013}]{2013MNRAS.436..430R}
{Roseboom} I.~G.,  et~al., 2013, \mn@doi [\mnras] {10.1093/mnras/stt1577},
  \href {http://adsabs.harvard.edu/abs/2013MNRAS.436..430R} {436, 430}

\bibitem[\protect\citeauthoryear{{Safarzadeh}, {Hayward}  \&
  {Ferguson}}{{Safarzadeh} et~al.}{2016}]{2016arXiv160407402S}
{Safarzadeh} M.,  {Hayward} C.~C.,   {Ferguson} H.~C.,  2016, preprint, \href
  {http://adsabs.harvard.edu/abs/2016arXiv160407402S} {} (\mn@eprint {arXiv}
  {1604.07402})

\bibitem[\protect\citeauthoryear{{Salmon} et~al.,}{{Salmon}
  et~al.}{2016}]{2016ApJ...827...20S}
{Salmon} B.,  et~al., 2016, \mn@doi [\apj] {10.3847/0004-637X/827/1/20}, \href
  {http://adsabs.harvard.edu/abs/2016ApJ...827...20S} {827, 20}

\bibitem[\protect\citeauthoryear{{Schreiber}, {Pannella}, {Leiton}, {Elbaz},
  {Wang}, {Okumura}  \& {Labb{\'e}}}{{Schreiber}
  et~al.}{2017}]{2017A&A...599A.134S}
{Schreiber} C.,  {Pannella} M.,  {Leiton} R.,  {Elbaz} D.,  {Wang} T.,
  {Okumura} K.,   {Labb{\'e}} I.,  2017, \mn@doi [\aap]
  {10.1051/0004-6361/201629155}, \href
  {http://adsabs.harvard.edu/abs/2017A%26A...599A.134S} {599, A134}

\bibitem[\protect\citeauthoryear{{S{\'e}rsic}}{{S{\'e}rsic}}{1963}]{1963BAAA....6...41S}
{S{\'e}rsic} J.~L.,  1963, Boletin de la Asociacion Argentina de Astronomia La
  Plata Argentina, \href {http://adsabs.harvard.edu/abs/1963BAAA....6...41S}
  {6, 41}

\bibitem[\protect\citeauthoryear{{Silva}, {Granato}, {Bressan}  \&
  {Danese}}{{Silva} et~al.}{1998}]{1998ApJ...509..103S}
{Silva} L.,  {Granato} G.~L.,  {Bressan} A.,   {Danese} L.,  1998, \mn@doi
  [\apj] {10.1086/306476}, \href
  {http://adsabs.harvard.edu/abs/1998ApJ...509..103S} {509, 103}

\bibitem[\protect\citeauthoryear{{Simpson} et~al.,}{{Simpson}
  et~al.}{2014}]{2014ApJ...788..125S}
{Simpson} J.~M.,  et~al., 2014, \mn@doi [\apj] {10.1088/0004-637X/788/2/125},
  \href {http://adsabs.harvard.edu/abs/2014ApJ...788..125S} {788, 125}

\bibitem[\protect\citeauthoryear{{Simpson} et~al.,}{{Simpson}
  et~al.}{2015}]{2015ApJ...807..128S}
{Simpson} J.~M.,  et~al., 2015, \mn@doi [\apj] {10.1088/0004-637X/807/2/128},
  \href {http://adsabs.harvard.edu/abs/2015ApJ...807..128S} {807, 128}

\bibitem[\protect\citeauthoryear{{Simpson} et~al.,}{{Simpson}
  et~al.}{2016}]{2016arXiv161103084S}
{Simpson} J.~M.,  et~al., 2016, preprint, \href
  {http://adsabs.harvard.edu/abs/2016arXiv161103084S} {} (\mn@eprint {arXiv}
  {1611.03084})

\bibitem[\protect\citeauthoryear{{Skelton} et~al.,}{{Skelton}
  et~al.}{2014}]{2014ApJS..214...24S}
{Skelton} R.~E.,  et~al., 2014, \mn@doi [\apjs] {10.1088/0067-0049/214/2/24},
  \href {http://adsabs.harvard.edu/abs/2014ApJS..214...24S} {214, 24}

\bibitem[\protect\citeauthoryear{{Smail}, {Ivison}  \& {Blain}}{{Smail}
  et~al.}{1997}]{1997ApJ...490L...5S}
{Smail} I.,  {Ivison} R.~J.,   {Blain} A.~W.,  1997, \mn@doi [\apjl]
  {10.1086/311017}, \href {http://adsabs.harvard.edu/abs/1997ApJ...490L...5S}
  {490, L5}

\bibitem[\protect\citeauthoryear{{Smail}, {Ivison}, {Blain}  \&
  {Kneib}}{{Smail} et~al.}{2002}]{2002MNRAS.331..495S}
{Smail} I.,  {Ivison} R.~J.,  {Blain} A.~W.,   {Kneib} J.-P.,  2002, \mn@doi
  [\mnras] {10.1046/j.1365-8711.2002.05203.x}, \href
  {http://adsabs.harvard.edu/abs/2002MNRAS.331..495S} {331, 495}

\bibitem[\protect\citeauthoryear{{Smol{\v c}i{\'c}} et~al.,}{{Smol{\v c}i{\'c}}
  et~al.}{2012a}]{2012ApJS..200...10S}
{Smol{\v c}i{\'c}} V.,  et~al., 2012a, \mn@doi [\apjs]
  {10.1088/0067-0049/200/1/10}, \href
  {http://adsabs.harvard.edu/abs/2012ApJS..200...10S} {200, 10}

\bibitem[\protect\citeauthoryear{{Smol{\v c}i{\'c}} et~al.,}{{Smol{\v c}i{\'c}}
  et~al.}{2012b}]{2012A&A...548A...4S}
{Smol{\v c}i{\'c}} V.,  et~al., 2012b, \mn@doi [\aap]
  {10.1051/0004-6361/201219368}, \href
  {http://adsabs.harvard.edu/abs/2012A%26A...548A...4S} {548, A4}

\bibitem[\protect\citeauthoryear{{Speagle}, {Steinhardt}, {Capak}  \&
  {Silverman}}{{Speagle} et~al.}{2014}]{2014ApJS..214...15S}
{Speagle} J.~S.,  {Steinhardt} C.~L.,  {Capak} P.~L.,   {Silverman} J.~D.,
  2014, \mn@doi [\apjs] {10.1088/0067-0049/214/2/15}, \href
  {http://adsabs.harvard.edu/abs/2014ApJS..214...15S} {214, 15}

\bibitem[\protect\citeauthoryear{{Swinbank}, {Chapman}, {Smail}, {Lindner},
  {Borys}, {Blain}, {Ivison}  \& {Lewis}}{{Swinbank}
  et~al.}{2006}]{2006MNRAS.371..465S}
{Swinbank} A.~M.,  {Chapman} S.~C.,  {Smail} I.,  {Lindner} C.,  {Borys} C.,
  {Blain} A.~W.,  {Ivison} R.~J.,   {Lewis} G.~F.,  2006, \mn@doi [\mnras]
  {10.1111/j.1365-2966.2006.10673.x}, \href
  {http://adsabs.harvard.edu/abs/2006MNRAS.371..465S} {371, 465}

\bibitem[\protect\citeauthoryear{{Swinbank} et~al.,}{{Swinbank}
  et~al.}{2014}]{2014MNRAS.438.1267S}
{Swinbank} A.~M.,  et~al., 2014, \mn@doi [\mnras] {10.1093/mnras/stt2273},
  \href {http://adsabs.harvard.edu/abs/2014MNRAS.438.1267S} {438, 1267}

\bibitem[\protect\citeauthoryear{{Symeonidis} et~al.,}{{Symeonidis}
  et~al.}{2013}]{2013MNRAS.431.2317S}
{Symeonidis} M.,  et~al., 2013, \mn@doi [\mnras] {10.1093/mnras/stt330}, \href
  {http://adsabs.harvard.edu/abs/2013MNRAS.431.2317S} {431, 2317}

\bibitem[\protect\citeauthoryear{{Tacconi} et~al.,}{{Tacconi}
  et~al.}{2013}]{2013ApJ...768...74T}
{Tacconi} L.~J.,  et~al., 2013, \mn@doi [\apj] {10.1088/0004-637X/768/1/74},
  \href {http://adsabs.harvard.edu/abs/2013ApJ...768...74T} {768, 74}

\bibitem[\protect\citeauthoryear{{Takagi}, {Hanami}  \& {Arimoto}}{{Takagi}
  et~al.}{2004}]{2004MNRAS.355..424T}
{Takagi} T.,  {Hanami} H.,   {Arimoto} N.,  2004, \mn@doi [\mnras]
  {10.1111/j.1365-2966.2004.08334.x}, \href
  {http://adsabs.harvard.edu/abs/2004MNRAS.355..424T} {355, 424}

\bibitem[\protect\citeauthoryear{{Takeuchi}, {Yuan}, {Ikeyama}, {Murata}  \&
  {Inoue}}{{Takeuchi} et~al.}{2012}]{2012ApJ...755..144T}
{Takeuchi} T.~T.,  {Yuan} F.-T.,  {Ikeyama} A.,  {Murata} K.~L.,   {Inoue}
  A.~K.,  2012, \mn@doi [\apj] {10.1088/0004-637X/755/2/144}, \href
  {http://adsabs.harvard.edu/abs/2012ApJ...755..144T} {755, 144}

\bibitem[\protect\citeauthoryear{{Targett} et~al.,}{{Targett}
  et~al.}{2013}]{2013MNRAS.432.2012T}
{Targett} T.~A.,  et~al., 2013, \mn@doi [\mnras] {10.1093/mnras/stt482}, \href
  {http://adsabs.harvard.edu/abs/2013MNRAS.432.2012T} {432, 2012}

\bibitem[\protect\citeauthoryear{{To}, {Wang}  \& {Owen}}{{To}
  et~al.}{2014}]{2014ApJ...792..139T}
{To} C.-H.,  {Wang} W.-H.,   {Owen} F.~N.,  2014, \mn@doi [\apj]
  {10.1088/0004-637X/792/2/139}, \href
  {http://adsabs.harvard.edu/abs/2014ApJ...792..139T} {792, 139}

\bibitem[\protect\citeauthoryear{{Toft} et~al.,}{{Toft}
  et~al.}{2014}]{2014ApJ...782...68T}
{Toft} S.,  et~al., 2014, \mn@doi [\apj] {10.1088/0004-637X/782/2/68}, \href
  {http://adsabs.harvard.edu/abs/2014ApJ...782...68T} {782, 68}

\bibitem[\protect\citeauthoryear{{Umehata} et~al.,}{{Umehata}
  et~al.}{2017}]{2017ApJ...835...98U}
{Umehata} H.,  et~al., 2017, \mn@doi [\apj] {10.3847/1538-4357/835/1/98}, \href
  {http://adsabs.harvard.edu/abs/2017ApJ...835...98U} {835, 98}

\bibitem[\protect\citeauthoryear{{Walter} et~al.,}{{Walter}
  et~al.}{2016}]{2016ApJ...833...67W}
{Walter} F.,  et~al., 2016, \mn@doi [\apj] {10.3847/1538-4357/833/1/67}, \href
  {http://adsabs.harvard.edu/abs/2016ApJ...833...67W} {833, 67}

\bibitem[\protect\citeauthoryear{{Wang}, {Cowie}, {Barger}  \&
  {Williams}}{{Wang} et~al.}{2011}]{2011ApJ...726L..18W}
{Wang} W.-H.,  {Cowie} L.~L.,  {Barger} A.~J.,   {Williams} J.~P.,  2011,
  \mn@doi [\apjl] {10.1088/2041-8205/726/2/L18}, \href
  {http://adsabs.harvard.edu/abs/2011ApJ...726L..18W} {726, L18}

\bibitem[\protect\citeauthoryear{{Wei{\ss}} et~al.,}{{Wei{\ss}}
  et~al.}{2009}]{2009ApJ...707.1201W}
{Wei{\ss}} A.,  et~al., 2009, \mn@doi [\apj] {10.1088/0004-637X/707/2/1201},
  \href {http://adsabs.harvard.edu/abs/2009ApJ...707.1201W} {707, 1201}

\bibitem[\protect\citeauthoryear{{Younger} et~al.,}{{Younger}
  et~al.}{2007}]{2007ApJ...671.1531Y}
{Younger} J.~D.,  et~al., 2007, \mn@doi [\apj] {10.1086/522776}, \href
  {http://adsabs.harvard.edu/abs/2007ApJ...671.1531Y} {671, 1531}

\bibitem[\protect\citeauthoryear{{Younger} et~al.,}{{Younger}
  et~al.}{2009}]{2009ApJ...704..803Y}
{Younger} J.~D.,  et~al., 2009, \mn@doi [\apj] {10.1088/0004-637X/704/1/803},
  \href {http://adsabs.harvard.edu/abs/2009ApJ...704..803Y} {704, 803}

\bibitem[\protect\citeauthoryear{{Yun} \& {Carilli}}{{Yun} \&
  {Carilli}}{2002}]{2002ApJ...568...88Y}
{Yun} M.~S.,  {Carilli} C.~L.,  2002, \mn@doi [\apj] {10.1086/338924}, \href
  {http://adsabs.harvard.edu/abs/2002ApJ...568...88Y} {568, 88}

\bibitem[\protect\citeauthoryear{{Yun} et~al.,}{{Yun}
  et~al.}{2008}]{2008MNRAS.389..333Y}
{Yun} M.~S.,  et~al., 2008, \mn@doi [\mnras]
  {10.1111/j.1365-2966.2008.13565.x}, \href
  {http://adsabs.harvard.edu/abs/2008MNRAS.389..333Y} {389, 333}

\bibitem[\protect\citeauthoryear{{Yun} et~al.,}{{Yun}
  et~al.}{2012}]{2012MNRAS.420..957Y}
{Yun} M.~S.,  et~al., 2012, \mn@doi [\mnras]
  {10.1111/j.1365-2966.2011.19898.x}, \href
  {http://adsabs.harvard.edu/abs/2012MNRAS.420..957Y} {420, 957}

\bibitem[\protect\citeauthoryear{{Zavala}, {Aretxaga}  \& {Hughes}}{{Zavala}
  et~al.}{2014}]{2014MNRAS.443.2384Z}
{Zavala} J.~A.,  {Aretxaga} I.,   {Hughes} D.~H.,  2014, \mn@doi [\mnras]
  {10.1093/mnras/stu1330}, \href
  {http://adsabs.harvard.edu/abs/2014MNRAS.443.2384Z} {443, 2384}

\bibitem[\protect\citeauthoryear{{Zavala} et~al.,}{{Zavala}
  et~al.}{2017}]{2017MNRAS.464.3369Z}
{Zavala} J.~A.,  et~al., 2017, \mn@doi [\mnras] {10.1093/mnras/stw2630}, \href
  {http://adsabs.harvard.edu/abs/2017MNRAS.464.3369Z} {464, 3369}

\bibitem[\protect\citeauthoryear{{Zavala} et~al.,}{{Zavala}
  et~al.}{2018}]{2018NatAs...2...56Z}
{Zavala} J.~A.,  et~al., 2018, \mn@doi [Nature Astronomy]
  {10.1038/s41550-017-0297-8}, \href
  {http://adsabs.harvard.edu/abs/2018NatAs...2...56Z} {2, 56}

\bibitem[\protect\citeauthoryear{{da Cunha}, {Charlot}  \& {Elbaz}}{{da Cunha}
  et~al.}{2008}]{2008MNRAS.388.1595D}
{da Cunha} E.,  {Charlot} S.,   {Elbaz} D.,  2008, \mn@doi [\mnras]
  {10.1111/j.1365-2966.2008.13535.x}, \href
  {http://adsabs.harvard.edu/abs/2008MNRAS.388.1595D} {388, 1595}

\bibitem[\protect\citeauthoryear{{da Cunha} et~al.,}{{da Cunha}
  et~al.}{2015}]{2015ApJ...806..110D}
{da Cunha} E.,  et~al., 2015, \mn@doi [\apj] {10.1088/0004-637X/806/1/110},
  \href {http://adsabs.harvard.edu/abs/2015ApJ...806..110D} {806, 110}

\bibitem[\protect\citeauthoryear{{van der Wel} et~al.,}{{van der Wel}
  et~al.}{2012}]{2012ApJS..203...24V}
{van der Wel} A.,  et~al., 2012, \mn@doi [\apjs] {10.1088/0067-0049/203/2/24},
  \href {http://adsabs.harvard.edu/abs/2012ApJS..203...24V} {203, 24}

\makeatother
\end{thebibliography}

% \begin{thebibliography}{99}
% \bibitem[\protect\citeauthoryear{Author}{2012}]{Author2012}
% Author A.~N., 2013, Journal of Improbable Astronomy, 1, 1
% \bibitem[\protect\citeauthoryear{Others}{2013}]{Others2013}
% Others S., 2012, Journal of Interesting Stuff, 17, 198
% \end{thebibliography}

%%%%%%%%%%%%%%%%%%%%%%%%%%%%%%%%%%%%%%%%%%%%%%%%%%

%%%%%%%%%%%%%%%%% APPENDICES %%%%%%%%%%%%%%%%%%%%%

\appendix

\section{Tables}

\begin{table*} \label{catalogue}
\caption{Counterpart identifications and physical properties of galaxies in our sample. The columns give: (1) \& (2) SCUBA-2 850 and 450\! $\micron$ source names, 
respectively; (3) \& (4) coordinates of the associated optical counterpart; (5) \& (6) redshift and the source where it comes from$^\dagger$; (7) stellar mass; (8) IR luminosity
(9) dust temperature, (10) UV star formation rate, and; (11)  UV spectral slope (from \citet{2014ApJS..214...24S}).
% and;  (10), (11) \& (12) the probabilities of being classified as a disk, spheroidal, or irregular, respectively (from \citealt{2015ApJS..221....8H}). 
The coordinates and flux densities at 450 and 850\ $\micron$ can be found in \citet{2017MNRAS.464.3369Z}.
}
\begin{center}
\begin{tabular}{ccccccccccc}
\hline
 850\! $\micron$-ID        &  450\! $\micron$-ID &    RA$_{\rm opt}$      &       DEC$_{\rm opt}$    &     $z$  & $z_s^\dagger$ &   log($M_*$) & log($L_{\rm FIR}$)  & $\rm T_d$ & SFR 1600\AA &$\beta$ \\ %  $f_{\rm disk}$ & $f_{\rm sph}$ & $f_{\rm irr}$\\ %UV_slope, morphology
                           &                     & [hh:mm:ss.s]           &[$^\circ$:`:"]           &           &            &   [$\rm M_\odot$] &  [$\rm L_\odot$] & [K] & [$\rm M_\odot/yr$] &  \\
\hline
  850.001 &  450.02     &    214.9109  &    52.9010  &   $2.240_{-0.003}^{+0.001}$ &  2      &    $ 11.10 \pm 0.07  $ &   $ 12.53 \pm   0.04$  & $  40.5  \pm  0.9$   & 5.5  &   -0.43   \\
  850.002 &  450.03     &    214.9144  &    52.8760  &   $0.73_{-0.00}^{+0.00}$    &  1      &    $ 10.87 \pm 0.10  $ &   $ 11.62 \pm   0.03$  & $  27.2  \pm  0.3$   & 30  &    1.11    \\
  850.003 &  450.05     &    214.9163  &    52.8913  &   $3.04_{-0.06}^{+0.00}$    &  2      &    $ 10.53 \pm 0.04  $ &   $ 12.76 \pm   0.04$  & $  52.7  \pm  1.5$   & 21.5  &   -0.88    \\
  850.004 &  450.25    &    214.9467  &    52.9101  &   $1.01_{-0.09}^{+0.08}$    &  3      &    $ 8.68  \pm  0.08  $ &              -         &              -       &  1.3 &    -1.49   \\
  850.005 &  450.08     &    214.9781  &    52.8115  &   $2.98_{-0.28}^{+0.28}$    &  3      &    $ 11.24  \pm 0.17    $ &   $ 12.87 \pm   0.05$  & $  57.4  \pm  1.5$   & -  &    -    \\
  850.006 &  450.13    &    214.9741  &    52.9060  &   $2.45_{-0.01}^{+0.12}$    &  2      &    $ 11.16 \pm 0.21  $ &   $ 12.67 \pm   0.04$  & $  57.1  \pm  1.2$   & 17    &   -1.18    \\
  850.007 &  450.20    &    214.9624  &    52.8703  &   $3.02_{-0.12}^{+0.13}$    &  3      &    $ 10.40 \pm 0.09  $ &   $ 12.20 \pm   0.11$  & $  39.1  \pm  4.2$   & 12.3  &   -2.11    \\
  850.008 &  450.15    &    215.0144  &    52.9013  &   $1.176_{-0.006}^{+0.004}$ &  2      &    $ 11.43 \pm 0.14  $ &   $ 12.21 \pm   0.04$  & $  46.3  \pm  0.8$   & 595   &   -0.93   \\
  850.009 &  450.12    &    214.8437  &    52.9110  &   $1.45_{-0.03}^{+0.03}$    &  3      &    $ 10.50 \pm 0.18  $ &   $ 12.07 \pm   0.05$  & $  36.2  \pm  0.8$   & 68.5  &   -0.72   \\
  850.010 &  450.14    &    214.9197  &    52.8397  &   $1.98_{-0.001}^{+0.29}$   &  2      &    $ 11.01 \pm 0.10  $ &   $ 12.14 \pm   0.07$  & $  37.6  \pm  1.5$   & 5.0  &    0.31   \\
  850.011 &  450.39    &    214.9224  &    52.8220  &   $3.26_{-0.10}^{+0.11}$    &  3      &    $ 10.49 \pm 0.42  $ &   $ 12.16 \pm   0.14$  & $  34.4  \pm  3.8$   & 124   &   -2.39    \\
  850.012 &  450.06     &    214.9462  &    52.8761  &   $0.35_{-0.05}^{+0.05}$    &  3      &    $ 9.80    \pm 0.04  $ &   $ 10.80 \pm   0.04$  & $  24.7  \pm  0.4$   &  3.4 &    -1.61   \\     
  850.013 &  450.07     &      -    &       -  &   $0.95_{-0.00}^{+0.10}$         &  4      &          - &              -         &              -       & -  &    -    \\
  850.014 &  450.46    &    214.8562  &    52.9287  &   $1.69_{-0.22}^{+0.22}$    &  3      &    $ 10.98  \pm 0.16    $ &   $ 11.95 \pm   0.08$  & $  32.3  \pm  1.4$   & -  &    -    \\    
  850.015 &  450.09     &    214.9381  &    52.8743  &   $1.56_{-0.01}^{+0.00}$    &  2      &    $ 10.71 \pm  0.05 $ &   $ 12.21 \pm   0.06$  & $  49.2  \pm  1.3$   & 5.1  &    0.17   \\
  850.016 &  450.56    &    215.0077  &    52.8271  &   $1.84_{-0.05}^{+0.05}$    &  3      &    $ 9.43   \pm  0.23   $ &   $ 12.23 \pm   0.08$  & $  38.5  \pm  2.3$   & -  &    -    \\  
  850.017 &  450.11    &    214.8988  &    52.8526  &   $2.37_{-0.02}^{+0.00}$    &  2      &    $ 11.26 \pm  0.29 $ &   $ 12.63 \pm   0.07$  & $  63.0  \pm  1.7$   & 19.8  &   -0.57   \\
  850.018 &  450.84    &    214.9748  &    52.8606  &   $3.70_{-0.16}^{+0.14}$    &  3      &    $ 10.35 \pm  0.40 $ &   $ 12.62 \pm   0.09$  & $  65.3  \pm  3.9$   & 0.1&    0.10  \\
  850.019 &  450.41    &    214.9710  &    52.9574  &   $2.353_{-0.003}^{+0.003}$ &  2      &    $ 10.67 \pm  0.05 $ &   $ 12.32 \pm   0.09$  & $  42.0  \pm  1.9$   & 23.8  &   -1.08    \\
  850.020 &  450.23    &    215.0310  &    52.9165  &   $1.087_{-0.004}^{+0.001}$ &  2      &    $ 10.55 \pm  0.08 $ &   $ 11.65 \pm   0.08$  & $  29.9  \pm  0.9$   & 2.9  &   -0.13   \\
  850.021 &  450.17    &    215.0542  &    52.9260  &   $1.037_{-0.002}^{+0.001}$ &  2      &    $ 10.58 \pm  0.15 $ &   $ 12.13 \pm   0.03$  & $  51.4  \pm  0.8$   & 0.4 &    1.63    \\
  850.022 &  450.01     &    214.9276  &    52.8480  &   $1.408_{-0.000}^{+0.046}$ &  2      &    $ 11.09 \pm  0.08 $ &   $ 12.50 \pm   0.03$  & $  61.9  \pm  0.9$   & 5.3  &   -1.85    \\
  850.023 &  450.65    &    -          &   -         &   $>3.95               $    &  4      &    -                   &              -         &              -       & -    &   -      \\
  850.024 &  450.04     &    214.9232  &    52.8822  &   $1.01_{-0.000}^{+0.000}$  &  1      &    $ 11.10 \pm  0.25  $ &   $ 11.95 \pm   0.05$  & $  42.7  \pm  0.8$   & 1.1  &   -0.26   \\
  850.025 &  450.29    &    214.8506  &    52.8664  &   $2.14_{-0.01}^{+0.03}$    &  2      &    $ 11.32 \pm  0.11 $ &   $ 11.88 \pm   0.13$  & $  31.7  \pm  3.4$   & 11.5  &    0.039  \\
  850.026 &  450.31    &     -     &     -    &   $2.25_{-0.30}^{+0.35}$         &  4       &          -             &              -         &              -       & -     &   -      \\
  850.028 &  450.69    &    214.8763  &    52.8521  &   $2.84_{-0.17}^{+0.02}$    &  2      &    $ 11.14 \pm  0.40 $ &   $ 12.37 \pm   0.11$  & $  49.3  \pm  3.0$   & 3.8  &   -2.14    \\
  850.029 &  450.21    &    214.8354  &    52.8438  &   $2.039_{-0.003}^{+0.001}$ &  2      &    $ 10.58 \pm  0.04 $ &   $ 12.50 \pm   0.06$  & $  60.9  \pm  1.6$   & 35.5  &   -0.88   \\
  850.030 &  450.10    &    214.8780  &    52.8768  &   $1.52_{-0.06}^{+0.06}$    &  3      &    $ 11.38 \pm  0.05 $ &   $ 11.87 \pm   0.09$  & $  33.9  \pm  2.0$   & 9.7  &   -0.19   \\
  850.031 &  450.49    &     -     &     -    &   $>3.65$                   &  4      &    - &              -         &              -       & -  &    -    \\
  850.032 &  450.63    &    214.9328  &    52.8330  &   $3.06_{-0.07}^{+0.02}$    &  2      &    $ 11.15 \pm  0.04 $ &   $ 12.40 \pm   0.11$  & $  51.5  \pm  3.4$   & 70    &   -0.81   \\
  850.033 &  450.59    &     -     &     -    &   $3.10_{-0.75}^{+2.10}$    &  4      &    - &              -         &              -       & -  &    -    \\
  850.034 &  450.66    &     -     &     -    &   $>3.6$                    &  4      &    - &              -         &              -       & -  &    -    \\
  850.037 &  450.64    &    215.0502  &    52.8539  &   $1.30_{-0.07}^{+0.07}$    &  3      &    $ 10.47  \pm   0.05  $ &   $ 11.85 \pm   0.11$  & $  33.8  \pm  2.0$   & -  &    -    \\       
  850.038 &  450.27    &    214.8647  &    52.8991  &   $1.93_{-0.01}^{+0.09}$    &  2      &    $ 10.92 \pm  0.14 $ &   $ 11.93 \pm   0.12$  & $  36.4  \pm  3.2$   & 7.4  &    0.17   \\
  850.039 &  450.40    &    215.0389  &    52.8544  &   $2.35_{-0.10}^{+0.10}$    &  3      &    $ 11.29  \pm  0.24   $ &   $ 12.43 \pm   0.10$  & $  56.5  \pm  2.5$   & -  &    -    \\        
  850.040 &  450.22    &     -     &     -    &   $3.55_{-0.90}^{+2.00}$    &  4      &    - &              -         &              -       & -  &    -    \\ 
  850.041 &  450.16    &    214.9928  &    52.8508  &   $1.68_{-0.06}^{+0.06}$    &  3      &    $ 10.56 \pm  0.05 $ &   $ 12.25 \pm   0.07$  & $  54.4  \pm  1.7$   & 2.2  &   -0.3     \\
  850.042 &  450.26    &    214.9800  &    52.9025  &   $2.03_{-0.02}^{+0.05}$    &  2      &    $ 11.03 \pm  0.14 $ &   $ 11.91 \pm   0.16$  & $  46.1  \pm  3.6$   & 2.0  &    1.63    \\
  850.043 &  450.73    &    214.9501  &    52.9383  &   $1.51_{-0.01}^{+0.01}$    &  2      &    $ 10.92 \pm  0.12 $ &   $ 11.88 \pm   0.11$  & $  40.1  \pm  1.8$   & 3.8  &    1.23    \\
  850.046 &  450.38    &     -     &     -    &   $1.60_{-0.25}^{+0.15}$    &  4      &    - &              -         &              -       & -  &    -    \\ 
  850.052 &  450.76    &    214.8278  &    52.9049  &   $1.70_{-0.03}^{+0.04}$    &  3      &    $ 11.12 \pm  0.05 $ &   $ 11.98 \pm   0.13$  & $  43.4  \pm  2.2$   & 3.6  &    0.07  \\
  850.059 &  450.24    &    214.8555  &    52.8489  &   $2.75_{-0.21}^{+0.24}$    &  3      &    $ 10.76 \pm  0.27 $ &   $ 12.30 \pm   0.16$  & $  53.2  \pm  4.6$   & 5.5  &    1.08    \\
  850.060 &  450.75    &    214.8495  &    52.9381  &   $0.94_{-0.06}^{+0.06}$    &  3      &    $ 11.08  \pm 0.13   $ &   $ 11.56 \pm   0.11$  & $  35.5  \pm  1.4$   & -  &    -    \\         
  850.065 &  450.18    &    214.8755  &    52.8665  &   $0.28_{-0.02}^{+0.02}$    &  3      &    $ 10.56 \pm  0.09 $ &   $ 10.84 \pm   0.05$  & $  30.7  \pm  0.6$   & 1.0  &    0.18   \\
  850.069 &  450.44    &    214.9548  &    52.8766  &   $0.51_{-0.00}^{+0.00}$    &  1      &    $ 11.21 \pm  0.06 $ &   $ 11.03 \pm   0.07$  & $  29.6  \pm  0.8$   & 6.1  &    1.39    \\
  850.070 &  450.34    &    214.9684  &    52.9251  &   $0.65_{-0.07}^{+0.01}$    &  2      &    $ 9.740  \pm 0.04 $ &   $ 10.99 \pm   0.14$  & $  26.8  \pm  1.4$   & 0.5 &    0.09  \\
  850.073 &  450.52    &    214.9480  &    52.8407  &   $1.358_{-0.003}^{+0.001}$ &  2      &    $ 11.24 \pm  0.17 $ &   $ 11.67 \pm   0.14$  & $  43.3  \pm  2.5$   & 0.1&    5.04    \\
  850.078 &  450.82    &    215.0377  &    52.8708  &   $0.56_{-0.04}^{+0.03}$    &  3      &    $ 10.65 \pm  0.16 $ &   $ 11.17 \pm   0.08$  & $  39.4  \pm  1.4$   & 0.4 &    1.63    \\
  850.079 &  450.54    &    215.0296  &    52.9364  &   $0.47_{-0.03}^{+0.03}$    &  3      &    $ 9.810  \pm 0.09 $ &   $ 10.41 \pm   0.20$  & $  20.2  \pm  1.5$   & 17.3  &   -1.03    \\
  850.085 &  450.33    &    215.0158  &    52.8569  &   $1.40_{-0.06}^{+0.07}$    &  3      &    $ 10.34  \pm   0.09  $ &   $ 11.64 \pm   0.19$  & $  39.3  \pm  3.1$   & -  &    -    \\                 
  850.092 &  450.60    &    214.8852  &    52.8158  &   $1.92_{-0.31}^{+0.01}$    &  2      &    $ 10.76 \pm  0.23 $ &   $ 11.77 \pm   0.47$  & $  43.3  \pm  8.2$   & 3.2  &   -0.06  \\
  850.095 &  450.37    &    214.9965  &    52.8141  &   $1.19_{-0.03}^{+0.03}$    &  3      &    $ 10.70  \pm   0.06   $ &   $ 11.66 \pm   0.14$  & $  49.2  \pm  3.0$   & -  &    -    \\       
  850.097 &  450.71    &    214.9206  &    52.8659  &   $1.01_{-0.000}^{+0.000}$  &  1      &    $ 11.33 \pm  0.32 $ &   $ 10.95 \pm   0.26$  & $  27.7  \pm  3.3$   & 2.3  &   -1.52    \\
  850.102 &  450.83    &     -     &     -    &   $2.35_{-0.95}^{+3.10}$    &  4      &    - &              -         &              -       & -  &    -    \\
  850.104 &  450.45    &    214.9866  &     52.8021 &   $ 2.5_{-0.21}^{+0.22}$    &  3      &    $ 10.83  \pm   0.06  $ &   $ 12.30 \pm   0.24$  & $  76.1  \pm  7.7$   & -  &    -    \\                           
 \hline                    
 \multicolumn{11}{l}{$^\dagger$(1) spectroscopic redshift; (2) HST grism redshift; (3) optical photometric redshift; (4) FIR photometric redshift (see \S\ref{z_dist_secc}).}\\

\end{tabular}                                                                                                        
\end{center}                                                                                                         
\end{table*}                                                                                                          

\begin{table*} 
{\bf Table A1.} (continuation)
\begin{center}
\begin{tabular}{ccccccccccc}
\hline
 850\! $\micron$-ID        &  450\! $\micron$-ID &    RA$_{\rm opt}$      &       DEC$_{\rm opt}$    &     $z$  &  $z_s$  &   $M_*$ &$L_{\rm FIR}$  & $\rm T_d$ & SFR1600 &$\beta$ \\ %  $f_{\rm disk}$ & $f_{\rm sph}$ & $f_{\rm irr}$\\ %UV_slope, morphology
                           &                     & [hh:mm:ss.s]           &[$^\circ$:`:"]           &           &            &   [$\rm M_\odot$] &  [$\rm L_\odot$] & [K] & [$\rm M_\odot/yr$] &  \\
\hline
  850.027 &    -       &      -    &      -   &   $ >5.3$                   &  4      &    - &              -         &              -       & -  &    -    \\
  850.035 &    -       &     -     &     -    &   $2.05_{-0.30}^{+0.90}$    &  4      &    - &              -         &              -       & -  &    -    \\
  850.036 &    -       &    -      &    -     &   $>3.85$                   &   4     &    - &              -         &              -       & -  &    -    \\
  850.044 &    -       &    215.0349  &    52.8914  &   $3.17_{-0.12}^{+0.14}$    &   3     &    $ 11.15 \pm  0.13 $ &   $ 12.40 \pm   0.12$  & $  57.0  \pm  4.5$   & 1.0  &    0.05   \\
  850.047 &    -       &    214.8598  &    52.8607  &   $3.79_{-0.53}^{+0.56}$    &   3     &    $ 10.97 \pm  0.10 $ &   $ 12.68 \pm   0.12$  & $  75.9  \pm  6.1$   & 4.2  &   -0.72   \\
  850.048 &    -       &    215.0454  &    52.8946  &   $1.89_{-0.01}^{+0.01}$    &   2     &    $ 11.35 \pm  0.06 $ &   $ 11.74 \pm   0.15$  & $  33.7  \pm  3.1$   & 4.3  &    0.15    \\
  850.049 &    -       &    -         &    -        &   $>4.75$                   &   4     &   -                    &              -         &              -       & -    &    -   \\
  850.050 &    -       &    214.9968  &    52.9408  &   $2.56_{-0.19}^{+0.41}$    &   3     &    $ 10.51 \pm  0.09 $ &   $ 12.23 \pm   0.13$  & $  52.6  \pm  3.4$   & 2.6  &    0.22   \\
  850.051 &    -       &    214.9951  &    52.9062  &   $2.78_{-0.06}^{+0.06}$    &   3     &    $ 11.09 \pm  0.19 $ &   $ 12.25 \pm   0.16$  & $  60.9  \pm  5.8$   & 19.4  &   -1.72    \\
  850.053 &    -       &    214.9935  &    52.8644  &   $2.99_{-0.16}^{+0.15}$    &   3     &    $ 10.68 \pm  0.40 $ &   $ 12.12 \pm   0.16$  & $  52.7  \pm  6.5$   & 0.1  &    1.42    \\
  850.054 &    -       &     214.8747 &     52.8438 &   $1.47_{-0.03}^{+0.03}$    &   3     &    $ 10.66 \pm  0.08 $ &   $ 11.73 \pm   0.14$  & $  33.2  \pm  2.0$   & 19.6  &   -0.58   \\
  850.055 &    -       &      -    &      -   &   $>4.5 $                   &   4     &   - &              -         &              -       & -  &    -    \\
  850.056 &    -       &     215.0273 &     52.8946 &   $2.48_{-0.01}^{+0.000}$   &   2     &    $ 10.64 \pm  0.08 $ &   $ 12.04 \pm   0.27$  & $  53.0  \pm  6.3$   & 4.52  &   -1.29    \\
  850.057 &    -       &     214.9969 &     52.8892 &   $3.69_{-0.04}^{+0.03}$    &   2     &    $ 10.28 \pm  0.06 $ &   $ 12.22 \pm   0.19$  & $  56.4  \pm  8.8$   & 101   &   -1.85    \\
  850.058 &    -       &      -    &      -   &   $>5.0$                    &   4     &    - &              -         &              -       & -  &    -    \\
  850.061 &    -       &      -    &      -   &   $2.70_{-0.65}^{+1.40}$    &   4     &    - &              -         &              -       & -  &    -    \\
  850.062 &    -       &     215.0618 &     52.9010 &   $2.18_{-0.62}^{+0.01}$    &   2     &    $ 10.88 \pm  0.13 $ &   $ 12.03 \pm   0.16$  & $  45.6  \pm  3.3$   & 0.9 &    1.03    \\
  850.063 &    -       &      -    &      -   &   $1.90_{-0.35}^{+0.65}$    &   4     &    - &              -         &              -       & -  &    -    \\ 
  850.064 &    -       &     -        &     -      &   $>4.4           $    &   4     &   -  &              -         &              -       & -   &   -    \\
  850.066 &    -       &     -        &     -      &   $>5.0           $    &   4     &    - &              -         &              -       & -  &    -    \\ 
  850.067 &    -       &     214.8293 &     52.8940 &   $2.95_{-0.10}^{+0.10}$    &   3     &    $ 11.40 \pm  0.68 $ &   $ 12.41 \pm   0.18$  & $  61.4  \pm  7.0$   & 5.7  &    1.22    \\
  850.068 &    -       &      -    &      -   &   $ >4.15$                  &   4     &    - &              -         &              -       & -  &    -    \\ 
  850.071 &    -       &      -    &      -   &   $2.70_{-0.75}^{+2.00}$    &   4     &    - &              -         &              -       & -  &    -    \\ 
  850.072 &    -       &     215.0159 &     52.9395 &   $3.36_{-0.33}^{+0.36}$    &   3     &    $ 10.61 \pm  0.81 $ &   $ 12.46 \pm   0.19$  & $  71.4  \pm  7.5$   & 5.7  &    0.75   \\
  850.074 &    -       &      -    &      -   &   $3.70_{-1.95}^{+2.05}$    &   4     &    - &              -         &              -       & -  &    -    \\ 
  850.075 &    -       &     214.8340 &     52.8701 &   $2.38_{-0.10}^{+0.10}$    &   3     &    $ 10.96 \pm  0.18 $ &   $ 12.17 \pm   0.18$  & $  57.6  \pm  4.5$   & 5.3  &    0.34   \\
  850.077 &    -       &     215.0297 &     52.8679 &   $3.48_{-0.52}^{+0.52}$    &   3     &    $ 11.08  \pm  0.10   $ &   $ 12.52 \pm   0.21$  & $  80.2  \pm  8.4$   & -  &    -    \\ 
  850.080 &    -       &     214.9231 &     52.9347 &   $2.92_{-0.05}^{+0.04}$    &   3     &    $ 10.38 \pm  0.07 $ &   $ 12.23 \pm   0.23$  & $  62.7  \pm  6.7$   & 44.9  &   -1.86    \\
  850.081 &    -       &     214.9546 &     52.8991 &   $1.73_{-0.06}^{+0.14}$    &   2     &    $ 10.32 \pm  0.16 $ &   $ 11.36 \pm   0.25$  & $  32.9  \pm  5.0$   & 9.6  &   -0.91   \\
  850.082 &    -       &     215.0557 &     52.8820 &   $2.81_{-0.09}^{+0.09}$    &   3     &    $ 10.77  \pm  0.06   $ &   $ 12.34 \pm   0.22$  & $  73.6  \pm  7.7$   & -  &    -    \\ 
  850.083 &    -       &     -     &     -    &   $2.55_{-0.85}^{+2.75}$    &   4     &    - &              -         &              -       & -  &    -    \\ 
   -      &   450.19  &     214.9169 &     52.8274 &   $1.228_{-0.002}^{+0.000}$ &   2     &    $ 11.23 \pm   0.16$ &   $ 11.89 \pm   0.07$  & $  43.8  \pm  1.4$   &  8.4 &     1.29   \\
   -      &   450.28  &     214.9459 &     52.8942 &   $1.210_{-0.003}^{+0.005}$ &   2     &    $ 10.92 \pm   0.06$ &   $ 11.51 \pm   0.24$  & $  35.9  \pm  3.2$   &  5.4 &    -1.89   \\
   -      &   450.35  &     214.9024 &     52.8637 &   $1.061_{-0.000}^{+0.000}$ &   1     &    $ 11.25 \pm   0.05$ &   $ 11.35 \pm   0.16$  & $  39.7  \pm  2.6$   &  5.2 &    -1.85   \\
   -      &   450.36  &     214.9361 &     52.8559 &   $1.455_{-0.005}^{+0.008}$ &   2     &    $ 10.63 \pm   0.10$ &   $ 11.56 \pm   0.19$  & $  40.8  \pm  3.8$   &  1.5  &    -0.19  \\
   -      &   450.42  &     214.9053 &     52.8510 &   $0.670_{-0.000}^{+0.000}$ &   1     &    $ 11.18 \pm   0.09$ &   $ 11.04 \pm   0.12$  & $  34.6  \pm  1.6$   &  24.8 &    -2.03   \\
   -      &   450.43  &     215.0204 &     52.9298 &   $2.59_{-0.18}^{+0.21}$    &   3     &    $ 10.31 \pm   0.13$ &   $ 11.52 \pm   0.55$  & $  35.7  \pm  3.4$   &  3.0 &    -1.29   \\
   -      &   450.47  &      -    &      -   &   $1.45_{-1.35}^{+2.60}$    &   4     &    - &              -         &              -       & -  &    -    \\ 
   -      &   450.48  &     214.9087 &     52.9279 &   $1.349_{-0.002}^{+0.002}$ &   2     &    $ 10.52 \pm   0.06$ &   $ 11.53 \pm   0.18$  & $  43.5  \pm  3.1$   &  11.5 &    -0.71  \\
 \hline                                                                                                               
\end{tabular}                                                                                                        
\end{center}                                                                                                         
\end{table*} 

\newpage
\section{Postage stamps}

\newpage

\begin{figure*}
\includegraphics[width=180mm]{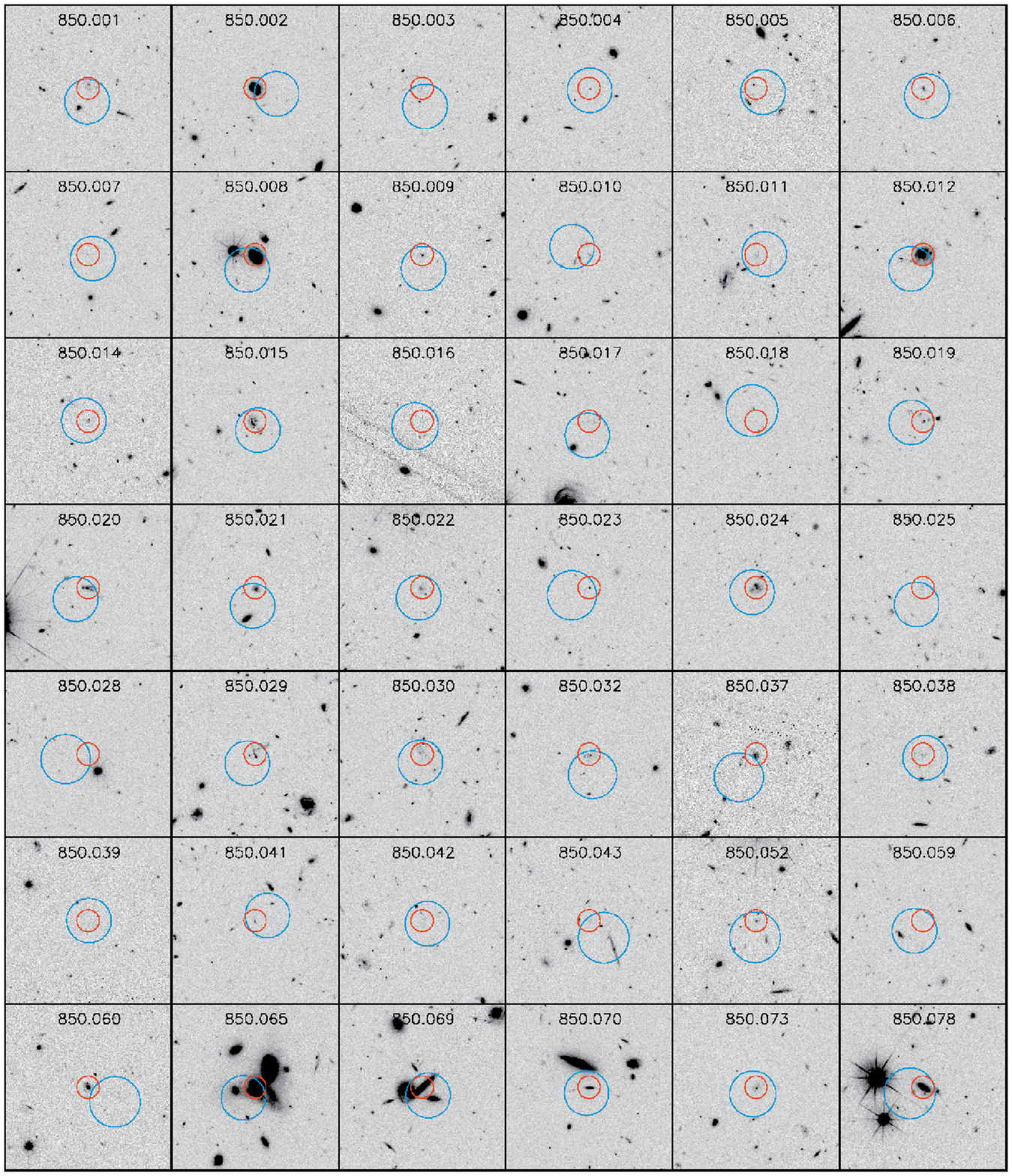}
\caption{30"\ $\times$\ 30" postage stamps in the HST/f814w band for our SMGs. The blue circle represents the variable search radius for counterparts
in the radio and IR wavelengths (centered at the SCUBA-2 position), and the red circle represents the final optical associated galaxy. The ID of each 
source is in the top of each panel.}
\label{postage_stamps}
\end{figure*}

\begin{figure*}
\includegraphics[width=180mm]{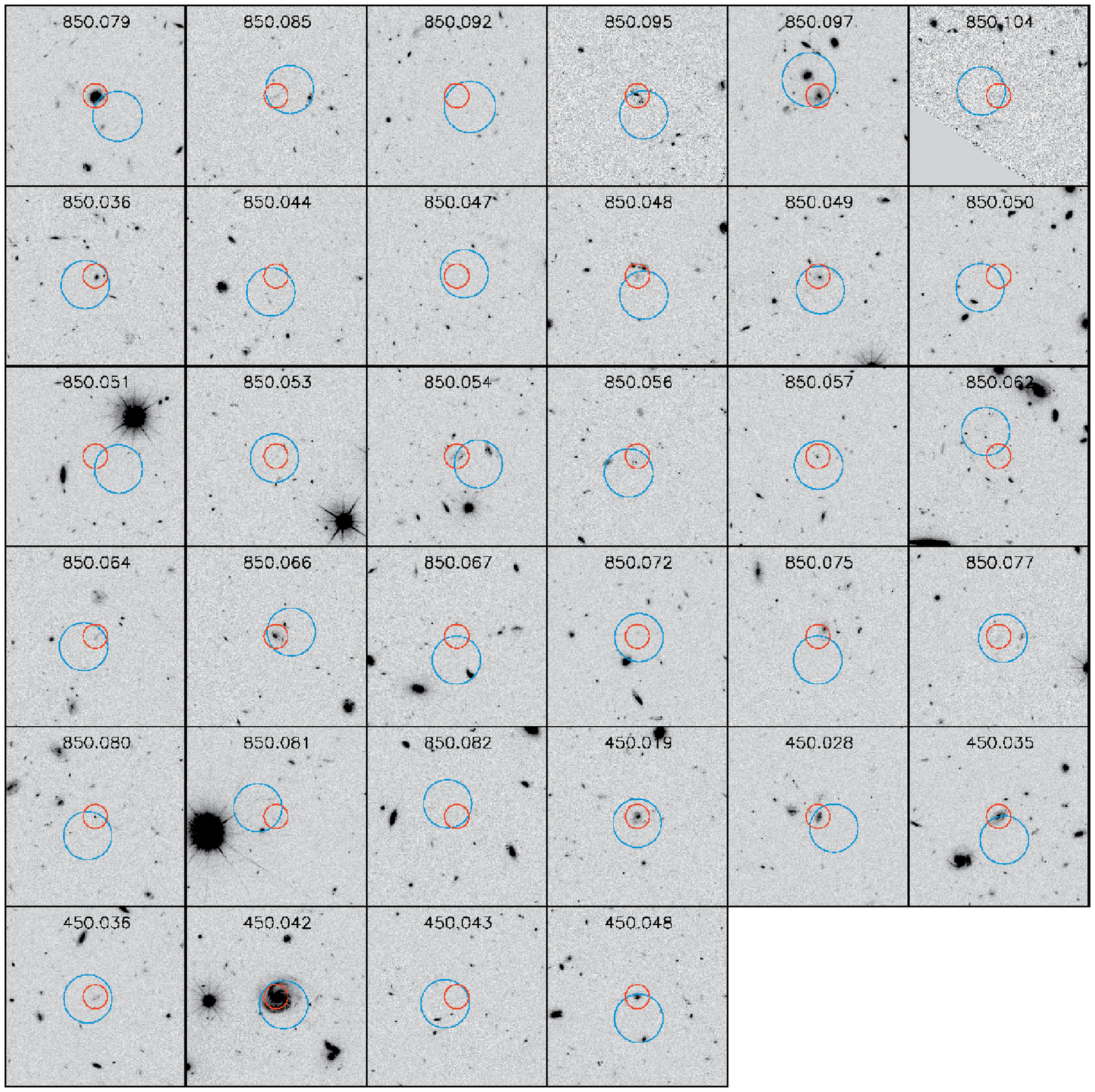}

\vspace{0.5cm}
{\bf Figure B1.} (continuation)
\end{figure*}

% Don't change these lines
\bsp	% typesetting comment
\label{lastpage}
\end{document}